\documentclass[aps,pre,twocolumn,notitlepage,floats,10pt]{revtex4-1}
\usepackage[hidelinks]{hyperref} 
\usepackage{amsmath}
\usepackage{lmodern}
\usepackage{amssymb}
\usepackage{epsfig}

\def\d{\mathrm{d}}
\def\ptop{{\vphantom\top}}
\newcommand{\be}{\begin{equation}}
\newcommand{\ee}{\end{equation}}
\newcommand{\beA}{\begin{equation}\begin{aligned}}
\newcommand{\eeA}{\end{aligned}\end{equation}}

\begin{document}
\title{Reservoir-induced stabilisation of a periodically driven classical spin
  chain:\\ local vs. global relaxation}
\date{\today}
\author{Thomas Veness}
\affiliation{
  School of Physics and Astronomy, University of Nottingham, Nottingham, 
  NG7 2RD, United Kingdom \\
  and 
  Centre for the Mathematics and Theoretical Physics of Quantum Non-equilibrium
  Systems, University of Nottingham, Nottingham NG7 2RD, United Kingdom
  }
\author{Kay Brandner}
\affiliation{
  School of Physics and Astronomy, University of Nottingham, Nottingham, 
  NG7 2RD, United Kingdom \\
  and 
  Centre for the Mathematics and Theoretical Physics of Quantum Non-equilibrium
  Systems, University of Nottingham, Nottingham NG7 2RD, United Kingdom
  }
\begin{abstract}
Floquet theory is an indispensable tool for analysing periodically-driven
quantum many-body systems. 
Although it does not universally extend to classical systems, some of its
methodologies can be adopted in the presence of well-separated timescales.
Here we use these tools to investigate the stroboscopic behaviours of a
classical spin chain that is driven by a periodic magnetic field and coupled to
a thermal reservoir. 
We detail and expand our previous work: 
we investigate the significance of higher-order corrections to the classical
Floquet-Magnus expansion in both the high- and low-frequency regimes;
explicitly probe the evolution the dynamics of the reservoir;
and further explore how the driven system synchronises with the applied field
at low frequencies.
In line with our earlier results, we find that the high-frequency regime is
characterised by a local Floquet-Gibbs ensemble with the reservoir acting as a
nearly-reversible heatsink.
At low frequencies, the driven system rapidly enters a synchronised state, which
can only be fully described in a global picture accounting for the concurrent
relaxation of the reservoir in a fictitious magnetic field arising from the
drive.
We highlight how the evolving nature of the reservoir may still be incorporated
in a local picture by introducing an effective temperature.
Finally, we argue that dissipative equations of motion for periodically-driven
many-body systems, at least at intermediate frequencies, must generically be
non-Markovian.
\end{abstract}

\maketitle

\begin{figure}[ht!]
  \includegraphics[width=.95\linewidth]{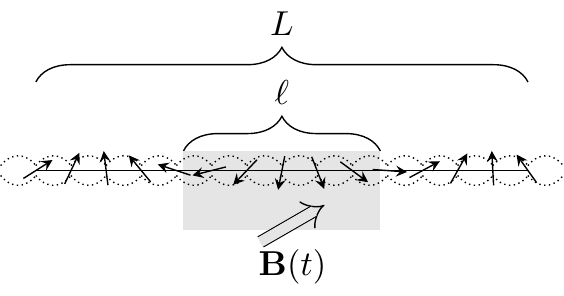}
  \includegraphics[width=0.95\linewidth]{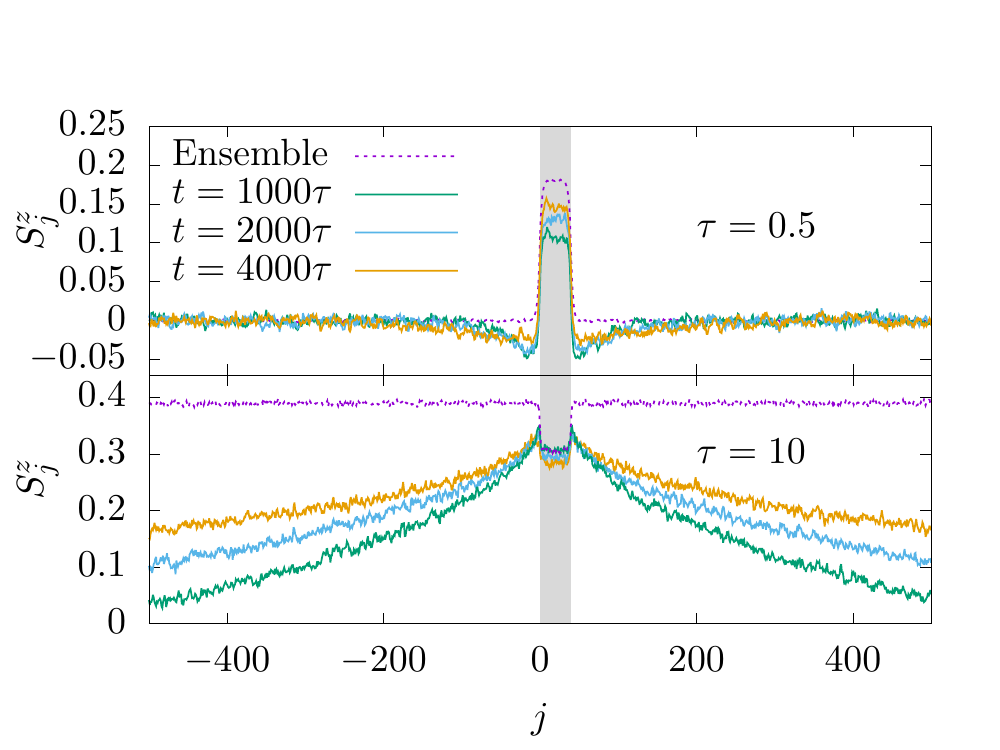}
  \caption{
  Above: System.
  A chain of $L$ classical spins, $l$ of which are subject to a uniform periodic
  magnetic field $\mathbf{B}(t)=\mathbf{B}(t+\tau)$.
  All spins have nearly isotropic nearest-neighbour interactions, with small
  disorder breaking conservation laws.
  Below: emerging states.
  The plots show a comparison between dynamical evolutions of the system over the
  time $t$ and predicted equilibrium expectation values of spatially resolved
  $z$-magnetisation.
  After a transient phase, both regimes may be described by effective Gibbs
  ensembles sampled via Monte Carlo techniques.
  For high frequencies, e.g. $\tau=0.5$, the driven sites (shaded region) locally
  see significant evolution.
  The reservoir, initialised in a thermal state with zero magnetisation, remains
  unchanged up to noise beyond a small correlation length.
  In contrast, the low-frequency profile, e.g. $\tau=10$, continues to spread deep
  into the reservoir at late times, indicating the global nature of the emerging
  steady state.
  We have set the parameters $L=2000$, $\ell=40$, $e_{\rm initial}=-0.66$,
  $\delta J=10^{-3}$, see Sec.~\ref{sec:system} for details.
  }
  \label{fig:sketch}
\end{figure}
\section{Introduction}
Floquet's theorem states that any homogeneous system of linear differential
equations with time-periodic coefficients can be mapped to an autonomous system
by means of a linear basis transformation \cite{Floquet-1883}.
This transformation carries the same periodicity as the original system an can
be chosen to become the identity at integer multiples of the period. 
When applied to the Schr\"odinger equation this theorem implies that any
periodically driven quantum system is connected by a time-periodic unitary
transformation to an undriven system whose dynamics is stroboscopically
equivalent \cite{Kitagawa-Oka-Brataas-Fu-Demler,Sambe,Torres-Kunold,
Bukov-DAlessio-Polkovnikov-review,Eckardt-rmp}.
Classical equations of motion, however, are generally non-linear and there is
indeed no intuitive counterpart to Floquet's theorem in Hamiltonian mechanics,
as can be seen from the following argument \cite{Anatoli-acknowledgement}.
Any autonomous classical system with one degree of freedom is integrable as
energy provides the required conserved quantity.
By extension, if it were always possible to find a time-periodic canonical
transformation that renders its Hamiltonian time-independent, any periodically
driven system with one degree of freedom would be integrable. 
This hypothesis is easily falsified by a counter-examples: Kapitza's pendulum,
for instance, is known to be non-integrable despite having only one degree of
freedom \cite{Arnold,Broer-Krauskopf,Landau-Lifshitz}.

This observation seems to leave us with a fundamental gap between quantum and
classical mechanics.
Indeed, driven by recent experimental advances \cite{KWW},
studies of periodically driven many-body systems have so far mainly focused on
the quantum regime, where Floquet theory has exposed a rich landscape of
phenomena including sharp notions of non-equilibrium phases with no static
counterpart \cite{Khemani-PRL,Curt-phase}, new perspectives on many-body quantum
chaos \cite{Chalker-PRX}, and the possibility of engineering specific band
structures through precisely tunable driving fields
\cite{Weitenberg-Simonet,Rudner}.
Still, although there is no `classical Floquet theorem', much of the methodology
used to describe periodically driven quantum systems, such as the Floquet-Magnus
expansion, formally extends to Hamiltonian systems.
It is therefore not a priori obvious what phenomenology is particular to the
quantum realm.

In fact, quantum and classical many-body systems are alike in that they
generically tend to absorb energy from a periodic drive until they approach a
trivial `infinite-temperature ensemble'
\cite{Ponte-Chandran-Papic-Abanin,Lazarides-heating,Russomanno-Silva-Santoro,Ikeda-Polkovnikov}.
Some routes to avoid this overheating, such as preventing thermalisation through
many-body localisation \cite{Curt-phase}, draw on quantum effects.
Others, like conservation laws \cite{Haldar-Das}, may cause observables to
synchronise thus giving rise to Gibbs ensembles with time-periodic Lagrange
multipliers \cite{Lazarides-Das-Moessner-periodic}; 
or the high-frequency limit, where heating rates are typically exponentially
suppressed in the driving frequency
\cite{Abanin-DeRoeck-Ho-Huveneers,Abanin-DeRoeck-Ho-Huveneers2,Abanin-DeRoeck-Huveneers,Mori-Kuwahara-Saito,Mori-Kuwahara-Saito2};
should however be equally accessible in the classical regime.

This paper represents a contribution to a growing literature oriented around
exploring the phenomenology of classical many-body systems
\cite{Hodson-Jarzynski,Howell,Nunnenkamp,McRoberts-Bilitewski-Haque-Moessner}. 
We aim to shed new light on the
so far relatively unexplored class of classical periodically driven many-body
systems, which is potentially ripe with interesting physics.
Furthermore, we set out to investigate how the dynamics of such a system can be
stabilised away from the high-frequency limit by coupling to a large thermal
reservoir.
To these ends, we consider a classical spin-chain with nearest-neighbour
interactions and weak disorder, a small fraction of which is subject to a
time-periodic magnetic field, with the remainder acting as a reservoir, see
Fig.~\ref{fig:sketch}.
Expanding on our previous work \cite{short-paper}, we simulate the full
dynamics of this system and compare its emergent steady states with those
predicted by Gibbs ensembles.

This analysis yields two major insights, which are exposed by the plots of
Fig.~\ref{fig:sketch}.
First, in the high-frequency regime, the driven part of the system quickly
settles to a stroboscopic steady state with residual heat uptake being
dissipated into the reservoir, which plays the r\^{o}le of a passive heat sink.
This steady state is well described by a Gibbs ensemble, whose temperature is
determined by the initial state of the reservoir.
The corresponding effective Hamiltonian is accurately determined by the lowest
orders of the classical Floquet-Magnus expansion, which is essentially a
systematic method to average over the periodic driving, order by order in the
inverse frequency.
Second, even at low frequencies, the driven part of the system quickly attains a
stroboscopic steady state, which survives well beyond initial transient behaviour
and is only slowly destabilised by residual heating.
Since, in contrast to the high-frequency regime, the driven spins can now follow
the applied magnetic field, this state is characterised by synchronisation with
the drive, which provides a new mechanism for the effective suppression of
energy absorption.
At the same time, the state of the reservoir is altered qualitatively as an
emerging magnetisation profile spreads out from the driven sites and eventually
covers the entire spin chain. 
This effect enables the redistribution of energy over large spatial distances
and can be understood as a relaxation process in a rotating reference frame.
In this picture, the entire system approaches a new Gibbs state, whose
Hamiltonian differs from the original one throughout the driven and the undriven
parts of the system.
As a result, the corresponding effective temperature can deviate substantially
from the initial temperature of the reservoir.

Thus, the two regimes are fundamentally different in nature.
At high frequencies, the driven system dissipatively relaxes as if being
locally quenched to a new Hamiltonian.
At low frequencies, the driven system rapidly synchronises with the external
magnetic field, while long-range correlations with the reservoir are established
and a new global Gibbs state is gradually approached.
Underpinned by profoundly different mechanisms, these two regimes are separated
only by a narrow crossover region in frequency space. 
This crossover features a rapid change in the energy absorption of the system,
and the scale for its onset is determined by the interaction strength between
neighbouring spins \cite{short-paper}.

Our analysis of this phenomenology proceeds as follows.
In Sec.~\ref{sec:system} we define the system and outline the numerical
techniques for both dynamical simulation, and for statistical sampling of
known distributions.
In Sec.~\ref{sec:undriven} we validate these numerical procedures for the
undriven system, where we already have a firm theoretical footing in the
standard results of statistical mechanics. 
Here, we set the scene for for the main quantities of interest.
In Sec.~\ref{sec:driven} we turn towards our main programme:
investigating the periodically-driven dynamics of a classical many-body system.
We approximately construct ensemble descriptions at both high and low frequencies,
going beyond the leading order analysis of Ref.~\cite{short-paper}.
We demonstrate that further corrections are indeed small and consistent with
observed data.
In Sec.~\ref{sec:reservoir} we investigate the behaviour of the reservoir
itself, its importance in establishing the synchronisation of observables in
the low-frequency ensemble, and comment on the non-Markovian nature of the
system.

\section{System and simulation} \label{sec:system}

\subsection{Dynamics}

The system under consideration, sketched in Fig.~\ref{fig:sketch}, comprises a
chain of $L$ three-component classical spins ${\bf S}_j$ normalised such that
$|{\bf S}_j| = 1$.
Its Hamiltonian is given by
\be
\mathcal{H}(t) = - \sum_{j=1}^L {\bf S}_j^\top J_j^\ptop {\bf S}_{j+1}^\ptop + \sum_{j=1}^\ell {\bf B}(t) \cdot {\bf S}_j,
\label{eq:Hfull}
\ee
where we assume periodic boundary conditions ${\bf S}_{L+1} = {\bf S}_1$.
The rotating magnetic field ${\bf B}(t) = \left(\cos (\omega t), \sin (\omega
    t), 0\right)^\top$, with period $\tau = 2\pi/\omega$, acts on the sites
$j=1,\ldots,\ell$, to which we refer as the `system proper'.
Accordingly, we call the undriven $L-\ell$ sites the `reservoir'.
The coupling matrices are diagonal $J_j = {\rm diag}[J_j^x,J_j^y,J_j^z]$.
To break any exact conservation laws constraining the dynamics, we choose the
$J_j^\alpha$ independently and identically distributed from a normal
distribution, with mean $J$ which we set equal to $1$ throughout, and variance
$\delta J$ i.e. $J_j^\alpha \sim \mathcal{N}(1,\delta J)$.
To match the dimensions of energies and frequencies, we formally set the reduced
Planck constant $\hbar$ equal to 1, hence the spins being dimensionless.

The equations of motion are determined by the Poisson bracket $\{
  S_j^\alpha, S_k^\beta \} = \delta_{jk} \varepsilon^{\alpha\beta\gamma}
  S_j^\gamma$ via Hamilton's equations
\be
\frac{\d {\bf S}_j}{\d t} = \{ \mathcal{H}(t), {\bf S}_j \}.
\ee
This rule yields a coupled system of non-linear differential equations for the spin
degrees of freedom,
\beA
\frac{\d {\bf S}_j}{\d t} &= - {\boldsymbol\Omega}_j \times {\bf S}_j, \\
{\boldsymbol\Omega}_j &= J_{j-1} {\bf S}_{j-1} + J_j {\bf S}_{j+1} -
\begin{cases}
  {\bf B}(t), & 1 \leq j \leq \ell, \\
  0,          & {\rm otherwise}\end{cases}
.
\label{eq:eom}
\eeA

If the applied field ${\bf B}$ were to vanish and all $J_j$ were the identity
matrix, the total magnetisation $\boldsymbol{\mathcal{M}} =
\frac{1}{L}\sum_{j=1}^L {\bf S}_j$ would be an exactly conserved quantity.
To apply the principles of statistical mechanics, we would then have to include
these additional conserved quantities through Lagrange multipliers determined by
the initial conditions, and any statistical sampling would have to respect this
constraint.
We add a small amount of Gaussian noise to the couplings $J_j$ to preclude such
fine-tuning. 
This disorder breaks all conservation laws, and even in the absence of driving
only energy conservation holds, but does not significantly affect the
macroscopic properties of the system otherwise.

To fully determine the dynamics, we now specify the initial conditions for the
equations of motion (\ref{eq:eom}).
To understand the general behaviour of the system, we take a statistical
approach rather than focusing on individual trajectories, which may be subject
to fluctuations.
Our initial states are therefore drawn from a Gibbs ensemble defined by
\beA
\mathcal{P}_0 &= e^{-\beta \mathcal{H}_0}/\mathcal{Z}_0, \\
\mathcal{H}_0 &= - \sum_{j=1}^L {\bf S}_j^\top J_j^\ptop {\bf S}_j^{\ptop} \label{eq:H0}
,
\eeA
with $\beta$ being an inverse temperature chosen to fix a specific initial mean
energy density, and $\mathcal{Z}_0$ being a normalisation constant.
With these initial conditions, the system would remain statistically invariant
under its time evolution if no magnetic field were applied.
Hence, when the field is applied to the system proper, the dynamics of the
reservoir are locally in equilibrium and we do not need to account for
quench-like effects.
We expect both the slow and fast driving regimes to result in minimal energy
absorption, and that the state of the reservoir is only gradually modified.
In the following, unless otherwise indicated, all presented results correspond to
averages over both initial conditions and realisations of disorder for $J_j$.

\subsection{Numerical techniques}\label{sec:numerical-simulation}
As given by Eq.~(\ref{eq:eom}), the instantaneous evolution of each spin ${\bf
S}_j$ is a rotation in an effective magnetic field determined by the on-site
magnetic field and the field from its nearest neighbours ${\bf S}_{j-1}$ and
${\bf S}_{j+1}$.
The dynamics of the spin-chain may therefore be efficiently simulated using
alternated updating \cite{Jin-et-al}.
The basic idea of this approach is to split the spin-chain into two interleaving
sub-chains $A$ and $B$ comprising the even and odd sites, respectively.
The local field for each spin in $A$ depends only on those in $B$, and vice
versa, and we can update these fields alternately.
More precisely, the technique we implement is drawn from
Refs.~\cite{Jin-et-al,Krech-Bunker-Landau} and uses the simplest
Suzuki-Trotter decomposition of the time-evolution operator from time $t$ to
$t+\delta t$, $U(t+\delta t, t)$.
That is, we have
\be
U(t+\delta t, t)
=
e^{\frac{\delta t}{2} \mathcal{L}^A_{t+\delta t/2}}
e^{\delta t \mathcal{L}^B_{t+\delta t/2}}
e^{\frac{\delta t}{2} \mathcal{L}^A_{t+\delta t/2}}
+\mathcal{O}(\delta t^3)
.
\ee
The Liouville operator $\mathcal{L}^{A/B}_{t+\delta t/2}$ generates rotations on
the sub-chain $A/B$ in the effective field determined from sub-chain $B/A$ at
time $t+\delta t/2$.
The error of this decomposition is bounded by terms of order $\delta t^3$.
As the propagation over a time step $\delta t$ is now formulated solely in
terms of rotations, the spin normalisation is manifestly preserved by this
procedure.

We will make extensive use of Monte Carlo (MC) techniques to sample from
particular Gibbs distributions.
We use standard Metropolis-Hastings sampling, where a site is chosen randomly
with equal probability, and a proposed update of the spin at the particular
site pointing in a new direction chosen uniformly from the surface of the unit
sphere.
The proposed update is then accepted if the energy of the new configuration
decreases i.e. $\Delta E \leq 0$, or accepted with probability 
$e^{-\beta \Delta E}$ if $\Delta E > 0$.
As the number of proposals accepted/rejected increases, this procedure is
asymptotically guaranteed to sample from the Gibbs ensemble 
\cite{Newman-Barkema}.

\section{Undriven system}\label{sec:undriven}
To set the stage for our main investigations and to confirm the validity of our
numerical approach, we first consider the undriven system, reproducing results
from Ref.~[\onlinecite{Jin-et-al}].
That is, we set ${\bf B} = 0$ so that the Hamiltonian of the entire spin chain
is given by $\mathcal{H}_0$ of Eq.~(\ref{eq:H0}). 
We initialise the entire spin chain in a random state with a fixed mean energy
and zero total magnetisation, which then evolves under the equations of motion
(\ref{eq:eom}).
The fundamental postulate of statistical mechanics asserts that time averages
and ensemble averages are equivalent in an ergodic system.
This equivalence extends to all moments of observables, and therefore their full
distributions.

In a canonical ensemble with inverse temperature $\beta$ the probability
distribution of any observable $\hat{O} = \hat{O}(\{ {\bf S}_j \})$ reads
\be
P_{\rm can}(O) = \frac{1}{\mathcal{Z}_0} \int \d{\bf S}_1 \ldots \d{\bf S}_L
e^{-\beta \mathcal{H}_0} \delta\left[O - \hat{O}(\{ {\bf S}_j \})\right],
\label{eq:canonP}
\ee
with $\mathcal{Z}_0$ being a normalisation, which sums the probability of all
configurations compatible with $\hat{O}=O$.
Note that we use the canonical ensemble as a matter of convenience throughout.
This approach should be equivalent to a microcanonical one up to finite-size
corrections.

To confirm that the undriven spin chain satisfies ergodicity, we compute the
full probability distributions of macroscopic observables over a \emph{single}
trajectory and compare them with the corresponding ensemble distribution
(\ref{eq:canonP}).
For an arbitrary observable $\hat{O}$, we may sample at times
$t_1,\ldots,t_\mathcal{N}$, and write the binned probability density with bin
width $\epsilon$ as
\be
P_{T}(O) = \frac{1}{\mathcal{N}} \sum_{n=1}^\mathcal{N}
\Pi_{\epsilon} \left[ O - \hat{O}( \{ {\bf S}_j(t_n) \} ) \right]
\label{eq:PT-def} ,
\ee
where the sample times are given by $t_n = t_0 + n \Delta t$, choosing $t_0$,
$\Delta t$, and $\mathcal{N}$ sufficiently large such that the result is
insensitive to specific values.
The function $\Pi_{\epsilon}$ counts the number of points along the
trajectory, where the value of $\hat{O}$ lies within the window of width
$\epsilon$ centred at $O$.
In other words, $\Pi_\epsilon$ is an approximation of a delta-function, formally
given by
\be
\Pi_\epsilon[x] = \frac{1}{\epsilon} 
\begin{cases}  1 & -\epsilon/2 \leq x \leq \epsilon/2 \\ 0 & {\rm otherwise}
\end{cases}
.
\ee
MC sampling is a numerical technique that generates samples in the correct
proportions as determined by a given probability distribution, without having
to exhaustively explore the phase space of the system.
We may therefore construct the ensemble distribution (\ref{eq:canonP}) for any
observable by using $M$ MC samples as outlined in
Sec.~\ref{sec:numerical-simulation}.
Specifically, the equivalent of Eq.~(\ref{eq:PT-def}) for the Gibbs ensemble is
\be
P_{{\rm can}}(O) = \frac{1}{M} \sum_{m=1}^M
\Pi_{\epsilon} \left[ O - \hat{O}( \{ {\bf S}_j^{(m)} \} ) \right]
,
\label{eq:PMC-def}
\ee
where ${\bf S}_j^{(m)}$ is the $m$th MC sample.
We note that the MC algorithm samples from the canonical ensemble, and due to
energy conservation the dynamical trajectory is sampling from the
microcanonical ensemble.
However, when comparing local observables involving only degrees of freedom of
the first $\ell\ll L$ sites, even the energy density will fluctuate, and the
equivalence of ensembles implies that all results should be identical up to
finite-size effects.

\begin{figure*}[!ht]
  \includegraphics[width=\linewidth]{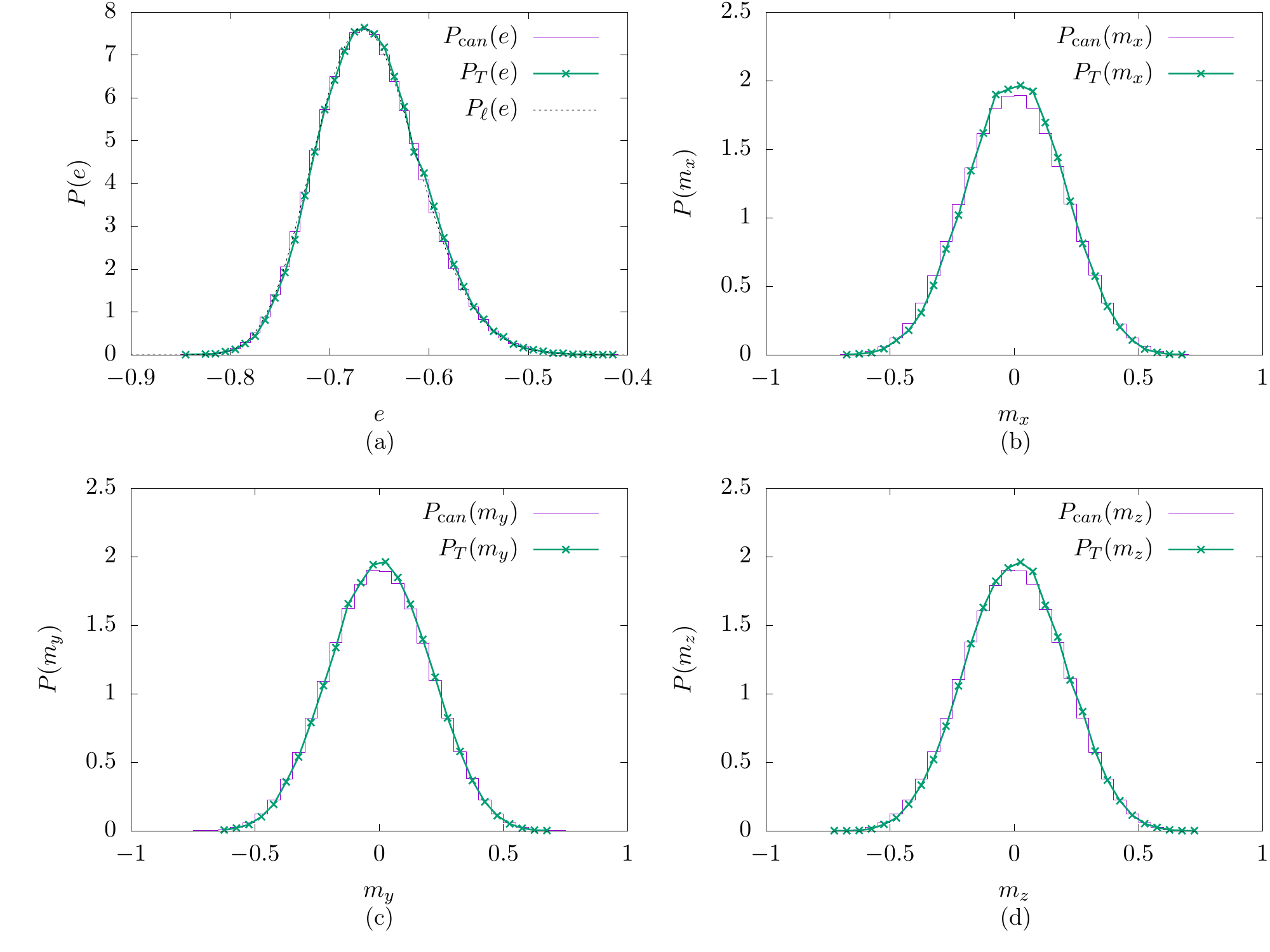}
  \caption{
    Undriven system. Histograms showing $P_T(O)$ of Eq.~(\ref{eq:PT-def})
    determined from one trajectory with $\mathcal{N}=64000$ samples (crosses),
    compared with $P_{\rm can}(O)$ of Eq.~(\ref{eq:PMC-def}), where each
    Monte-Carlo curve is binned from $M=64000$ samples.
    The shown observables are
    (a) energy density,
    (b) $x$-magnetisation,
    (c) $y$-magnetisation,
    (d) $z$-magnetisation.
    For all plots, we have set $L=2000$, $\ell=40$, 
    $\delta J = 2\times10^{-3}$, and chosen the energy density of the entire system
    $e_{\rm initial}=-0.66$, which implicitly fixes the inverse temperature $\beta$
    for the MC simulation.
    The exact result (dashed line) of Eq.~(\ref{eq:Pe-finite-size}) is also
    shown in (a).
  } \label{fig:mchist}
\end{figure*}

In Fig.~\ref{fig:mchist}, we show that single-trajectory and ensemble
distributions are in excellent agreement for the representative observables
\beA
{\bf m} &= \frac{1}{\ell} \sum_{j=1}^\ell {\bf S}_j, \\
e &= -\frac{1}{\ell-1} \sum_{j=1}^{\ell-1} {\bf S}_j^\top J^\ptop_j {\bf S}^\ptop_{j+1},
\label{eq:el}
\eeA
i.e. the magnetisation and energy density of the system proper.
This result confirms that the undriven system behaves ergodically even if
$\delta J = 10^{-3}$ is small.

For the isotropic chain, $\delta J = 0$, we may also compute the exact
probability distribution for the energy density at large $\ell$.
We here consider open boundary conditions, which are equivalent to periodic
boundary conditions up to finite-size corrections.
Firstly, the partition function is given by
\be
Z = \int \d{\bf S}_1 \cdots \d{\bf S}_\ell\,
e^{\beta \sum_{j=1}^{\ell-1} {\bf S}_j \cdot {\bf S}_{j+1}}
,
\ee
where $\d {\bf S}_j = \frac{1}{4\pi} \sin \theta_j \d \theta_j \d \phi_j$
indicates integration over the spherical angles of each spin.
Rewriting this expression in terms of the angle between adjacent spins in the
chain we find
\be
Z =\left[ \frac{1}{2} \int_{-1}^1 \d \cos \theta e^{\beta \cos\theta} \right]^{\ell-1}
= \left( \frac{\sinh \beta }{\beta } \right)^{\ell-1}
.
\ee
Hence, the free energy density of the isotropic chain at inverse temperature
$\beta$ is given by \cite{Parsons}
\be
f(\beta) = -\frac{1}{\beta(\ell-1)} \ln Z =
-\frac{1}{\beta} \ln \frac{\sinh \beta }{\beta }.
\ee
We may thus calculate the probability distribution for the energy density $e$
of Eq.~(\ref{eq:el}) as
\beA
P(e) &= \frac{1}{Z} \int \d{\bf S}_1 \cdots \d{\bf S}_{\ell} e^{\beta \sum_{j=1}^{\ell-1} {\bf S}_j \cdot {\bf S} _{j+1}}\\
& \int_{-\infty}^\infty \frac{\d z}{2\pi}e^{iz\left( e(\ell-1) + \sum_{j=1}^{\ell-1} {\bf S}_j \cdot {\bf S}_{j+1} \right)},
\eeA
where we have used the Fourier representation of the Dirac delta function
appearing in Eq.~(\ref{eq:canonP}).
Employing the method of steepest descent \cite{erdelyi}, this integral may be
approximated as
\be
P_\ell(e) = \frac{\mathcal{Q}}{\sqrt{(\ell-1) |g''(z^*)|}} e^{-(\ell-1) g(z^*)},
\label{eq:Pe-finite-size}
\ee
where $\mathcal{Q}$ is a normalisation constant,
and $\beta$ is chosen to fix the mean energy density 
$\frac{1}{\beta} - \coth \beta = \langle e \rangle$.
The function $g(z) = (\beta - z)f(\beta-z) + e z$ determines the saddle-point,
which is given by $z^*$ such that $g'(z^*) = 0$.
Fig.~\ref{fig:mchist}(a) shows that the exact distribution
(\ref{eq:Pe-finite-size}) agrees well with both numerical approaches.
This result confirms weak disorder $\delta J \ll 1$ eliminates conservation
laws but has no visible effect on the distributions of macroscopic observables.

\section{Driven system}\label{sec:driven}

Having established that the undriven system is ergodic, we now turn to the
effect of periodic driving.
Our basic expectations are as follows:
at high frequencies, the driving `averages out' and the dynamics are described
by the Hamiltonian with no applied field $\mathcal{H}_0$ of Eq.~(\ref{eq:H0});
at very low frequencies, the system relaxes to the instantaneous Hamiltonian.
Here, we develop these expectations into quantitative descriptions of the two
regimes, which we show to predict the behaviour of the system over a remarkably
wide range of frequencies \cite{short-paper}.

\subsection{High frequency} \label{sec:perturbations}

Intuitively, the physics at high frequencies should be well approximated by the
averaged Hamiltonian.
This intuition is captured by the Floquet-Magnus expansion.
This technique proceeds by assuming that a time-independent generator for the
stroboscopic dynamics does exist, and then constructs it order-by-order in the
driving period \cite{Blanes-Casas-Oteo-Ros}.
The Floquet-Magnus expansion was originally devised for linear systems, and is
expansively used for the description of periodically-driven quantum systems.
It can, however, be adapted to classical Hamiltonian systems, whose dynamics
are non-linear, by formally treating the Poisson bracket as a linear operator.
That is, we write the equations of motion for a phase-space observable of the
isolated system proper,
$\hat{O}({\bf S}_1,\ldots, {\bf S}_\ell)$,
in the form
\be
\frac{\d \hat{O}(t)}{\d t} = \mathcal{L}_{H(t)} \hat{O}(t),
\label{eq:liouville}
\ee
where $\mathcal{L}_{H(t)} \cdot = \{ H(t), \cdot \}$ denotes the Liouville
operator and
\be
H(t)  = -\sum_{j=1}^{\ell -1} {\bf S}_j^\top J^\ptop_j {\bf S}_{j+1}^\ptop
+ \sum_{j=1}^\ell {\bf B}(t) \cdot {\bf S}_j
\ee
is the Hamiltonian of the driven system proper.
If the Floquet theorem were to hold, the formal solution of
Eq.~(\ref{eq:liouville}) at $t=n\tau$ would take the form
\be
\hat{O}(n\tau)= e^{n\tau \mathcal{L}_{H_F}} \hat{O}(0),
\ee
where $H_F$ would be the Floquet Hamiltonian.
While such an object does not exist in general, it can still be perturbative
constructed close to the infinite frequency limit $\tau\to 0$, where 
$H_F \to \frac{1}{\tau} \int_0^\tau \d t H(t)$.
To this end, we introduce the dimensionless time $s = t/\tau$ and make the
ansatz
\be
\hat{O}'(s) = \hat{O}(s\tau) = e^{\mathcal{L}_{K(s)} } \hat{O}'(s),
\ee
with $0\leq s\leq 1$ and the Magnus Hamiltonian
\be
K(s) = \sum_{n=0}^\infty \tau^n K^{(n)}(s).
\ee
Upon inserting this ansatz into the equation of motion for $\hat{O}'(s)$ and
following the formal steps of the conventional Magnus expansion
\cite{Kitagawa-Oka-Brataas-Fu-Demler}, $K(s)$ can be determined order by order.
Returning to original units then yields the effective Floquet Hamiltonian
\be
H_F = \frac{1}{\tau} K(t=\tau) = \sum_{n=0}^\infty H_F^{(n)},
\label{eq:HF}
\ee
where the first few terms are given by
\beA
H_F^{(0)} &= \frac{1}{\tau} \int_0^\tau \d t\, H(t),\\
H_F^{(1)} &= \frac{1}{2! \tau} \int_0^\tau \d t_1 \int_0^{t_1} \d t_2\, \{ H(t_1), H(t_2) \} ,\\
H_F^{(2)} &= \frac{1}{3! \tau} \int_0^\tau \d t_1 \int_0^{t_1} \d t_2 \int_0^{t_2} \d t_3\, \\
&
\left\{ H(t_1), \{ H(t_2), H(t_3) \} \right\}\\
&
+ \left\{ H(t_3), \{ H(t_2), H(t_1) \} \right\} .
\eeA
This expansion can in general only be expected to be asymptotic
\cite{Bender-Orszag,Berry,mori-floquet-prethermal}, and one usually truncates
the sum in Eq.~(\ref{eq:HF}) after the first few terms.

Evaluating the first two expressions explicitly, we find
\beA
H_F^{(0)} &= -\sum_{j=1}^L {\bf S}^\top_j J_j^\ptop {\bf S}^\ptop_{j+1},\\
H_F^{(1)} &= -\frac{1}{\omega}\sum_{j=1}^\ell
\left(
\frac{1}{2}S_j^z
+\hat{\bf y}\cdot ({\boldsymbol\Omega_j}\times {\bf S}_j)
\right).
\label{eq:HF0-HF1}
\eeA
Noting that $H_F^{(0)}$ is equal to the Hamiltonian of the initial ensemble
Eq.~(\ref{eq:H0}), we must consider at least the first order in $\tau$ of the
Floquet-Magnus expansion to observe deviations from the initial state.
In our previous work the corrections from $H_F^{(1)}$ were observable and
consistent with the true dynamics for $\tau \lesssim 1$ \cite{short-paper}.
For clarity, we may write
\be
H_F^{(n)} = h_{\rm SB}^{(n)} + h_{\rm MB}^{(n)},
\ee
where $h_{\rm SB}^{(n)}$ accounts for single body terms, and $h_{\rm MB}^{(n)}$
accounts for many-body terms.
Accordingly,
\beA
h_{\rm SB}^{(1)} &= -\frac{1}{2\omega}\sum_{j=1}^\ell
\frac{1}{2}S_j^z ,
\\
h_{\rm MB}^{(1)} &= -\frac{\hat{\bf y}}{\omega} \cdot\sum_{j=1}^\ell
{\boldsymbol\Omega_j}\times {\bf S}_j.
\eeA

It is interesting to ask if including higher-order corrections from $H_F^{(2)}$
improves the results of Ref.~\cite{short-paper} further.
Calculating $H_F^{(2)} = h_{\rm SB}^{(2)} + h_{\rm MB}^{(2)}$ yields
\beA
h_{\rm SB}^{(2)} &= \frac{1}{2\omega^2} \sum_{j=1}^\ell S_j^x, \\
h_{\rm MB}^{(2)} &=
\frac{1}{2\omega^2} \sum_{j=1}^\ell
\left\{
2\hat{\bf x} \cdot \left[
  {\boldsymbol\Omega}_j \times \left( {\boldsymbol\Omega}_j \times {\bf S}_j \right)
  \right.\right.\\
  &\left.\left. +
  \left( J_{j-1} \left( {\boldsymbol\Omega}_{j-1} \times {\bf S}_{j-1} \right)
  + J_j\left( {\boldsymbol\Omega}_{j+1} \times {\bf S}_{j+1} \right)\right)\times {\bf S}_j
  \right]\right.\\
&\left.+\frac{1}{2} \left(
3 \Omega^x_j S_j^x + \Omega_j^y S_j^y
- J_{j-1}^y S_j^z S_{j-1}^z
\right. \right.\\
&\left.\left.
- J_j^y S_j^z S_{j+1}^z
- J_{j-1}^z S_j^y S_{j-1}^y
- J_j^z S_j^y S_{j+1}^y
\right. \right.\\
&\left. \left.
-3 J_{j-1}^x S_j^z S_{j-1}^z
-3 J_j^x S_j^z S_{j+1}^z
\right.\right.\\
&\left.\left.
-3 J_{j-1}^z S_j^x S_{j-1}^x
-3 J_j^z S_j^x S_{j+1}^x
\right)
\right\}
.
\eeA
The single-body terms $h_{\rm SB}^{(n)}$ constitute the Floquet-Magnus
expansion under the free Hamiltonian $H_{\rm free}= \sum_{j=1}^\ell {\bf
B}(t)\cdot {\bf S}_j$, a linear system for which the expansion may be resummed
to yield the effective field of the rotating frame ${\bf B}_{\rm rot} = \hat{\bf
x} - \omega \hat{\bf z}$ \cite{feldman}.
The many-body terms have more complex structure and become increasingly
non-local at each subsequent order.

\begin{figure*}[!ht]
  \includegraphics[width=.95\linewidth]{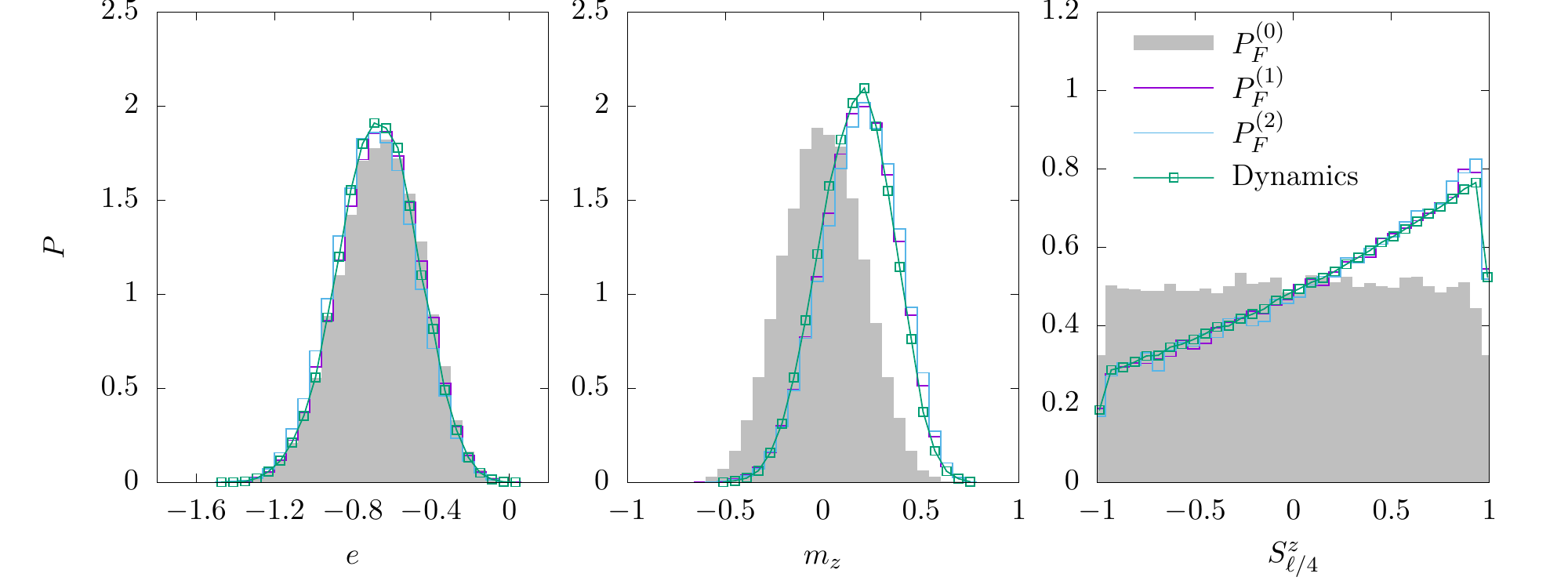}
  \caption{
  High-frequency picture.
  Histogram distributions for $\tau = 0.5$ of the energy density and
  $z$-magnetisation of the system proper, $e$ and $m_z$, and the
  $z$-component of the spin at site $f=\ell/4$, $S^z_{\ell/4}$, in the
  $N$th order Floquet-Magnus ensembles as defined by Eq.~(\ref{eq:PF-N})
  for $N=0,1,2$ compared with dynamical evaluations.
  Statistical ensembles are sampled according to
  Eq.~(\ref{eq:PMC-def}), and the dynamical ones according to
  Eq.~(\ref{eq:PT-def}) using multiple trajectories and choosing $t_0$
  long enough for observables to relax.
  For all plots, we have set $L=2000$, $\ell=40$, $\delta J = 0.02$, and
  chosen the initial energy density $e_{\rm initial}=-0.66$, which fixes
  the inverse temperature $\beta$ for the MC simulation.
  Due to the minimal energy absorption, we sample multiple points from
  each trajectory after a transient period to improve statistics.
  Here we sample 4000 points each from 200 separate trajectories, from the
  interval $t=6000\tau$ to $t=10,000\tau$.
  }
  \label{fig:high-freq-2nd-order}
\end{figure*}

Assuming that the stroboscopic dynamics of the system proper is asymptotically
equivalent to the autonomous dynamics of the associated Floquet system, we may
define the $N$th order Floquet-Magnus ensemble by
\be
P_F^{(N)} = \frac{1}{Z_F^{(N)}} \exp\left(-\beta \sum_{n=0}^N H_F^{(n)}\right),
\label{eq:PF-N}
\ee
where $Z_F^{(N)}$ accounts for normalisation.
As we expect absorption to be exponentially small in $\omega$ at high
frequencies \cite{rubio-abadal}, we neglect heating and assume the temperature
of the ensemble to be well approximated by that of the initial state defined by
Eq.~(\ref{eq:H0}).
As may be seen in Fig.~\ref{fig:high-freq-2nd-order}, including additional terms
in the Magnus ensemble beyond $N=1$ produces only minimal improvement on the
previously observed results at $\tau = 0.5$.
We can thus conclude that the two lowest orders of the Floquet-Magnus expansion
are indeed sufficient to describe the stroboscopic steady state of the system
proper, even well away from the infinite-frequency limit.

\subsection{Low frequency} \label{sec:low-freq}

\subsubsection{Leading order}

The results of the previous section show that the high-frequency regime can be
fully described in a local picture focusing only on the system proper and
treating the reservoir as a passive heat sink.
To understand the low-frequency regime, we must adopt a global picture, where
both system proper and reservoir are affected by the driving.  This approach is
motivated by the observation that the stroboscopic dynamics of the full system
are nearly equivalent to the dynamics of an autonomous Floquet system for
sufficiently small $\delta J$ but arbitrary $\omega$.
Specifically, upon introducing the rotating-frame variables
\be
\widetilde{\bf S}_j = \begin{pmatrix} \cos (\omega t) & \sin (\omega t) & 0 \\ -\sin (\omega t) & \cos (\omega t) & 0 \\ 0 & 0 & 1 \end{pmatrix} {\bf S}_j,
\ee
the equations of motion (\ref{eq:eom}) may be recast as
\beA
\frac{\d \widetilde{\bf S}_j}{\d t} &= -\widetilde{\boldsymbol\Omega}_j \times \widetilde{\bf S}_j + \mathcal{O}(\delta J),\\
\widetilde{\boldsymbol\Omega}_j &= -(\bar{J}_{j-1}\widetilde{\bf S}_{j-1} + \bar{J}_j\widetilde{\bf S}_{j+1})\\
& - \omega \hat{\bf z} + \begin{cases} \hat{\bf x},& 1\leq j \leq \ell,\\ 0,& {\rm otherwise} \end{cases}.
\label{eq:rotdrop}
\eeA
Thus, upon neglecting time-dependent terms of order $\delta J$, the dynamics in
the rotating frame are generated by the time-independent Hamiltonian
\be
\mathcal{H}_{\rm rot} = - \sum_{j=1}^{L} \left( \widetilde{\bf S}_j^\top \bar{J}_j^\ptop \widetilde{\bf S}_{j+1}^\ptop +\omega \hat{\bf z} \cdot \widetilde{\bf S}_j \right) + \sum_{j=1}^\ell \hat{\bf x} \cdot \widetilde{\bf S}_j,
\label{eq:Hrot-L}
\ee
where $\bar{J}_j = {\rm diag}\left[ \frac{1}{2} \left( J_j^x + J_j^y\right),
\frac{1}{2}  \left( J_j^x + J_j^y \right), J_j^z \right]$.

This result shows that the observable $\mathcal{H}_{\rm rot}$, which plays the
r\^{o}le of a global Floquet Hamiltonian, is nearly conserved in the rotating
frame. At the same time $\mathcal{H}_{\rm rot}$ is a sum of local densities.
Hence, by usual arguments of statistical mechanics, the entire system should
relax to an ensemble of the form
\be
\mathcal{P}_{\rm rot} = e^{-\beta_{\rm rot} \mathcal{H}_{\rm
      rot}}/\mathcal{Z}_{\rm rot}. \label{eq:rotL}
\ee
Note that formally, the system should be described by a microcanonical
ensemble.
However, for all local observables the canonical ensemble above should provide
an equivalent description up to finite size effects.

The effective inverse temperature $\beta_{\rm rot}$ may now be determined from
the fact that $\mathcal{H}_{\rm rot}$ is an almost conserved quantity:
assuming that the system has fully relaxed to the ensemble described by
Eq.~(\ref{eq:rotL}), and that persistent heating has only minor effects, the
mean value of $\mathcal{H}_{\rm rot}$ in the initial state $\mathcal{P}_0$ of
Eq.~(\ref{eq:H0}) should be nearly identical to its mean value in the steady
state of Eq.~(\ref{eq:rotL}).
That is, the equation
\be
\int \d{\bf S}_1 \cdots \d{\bf S}_L \mathcal{H}_{\rm rot} \mathcal{P}_0
=
\int \d{\bf S}_1 \cdots \d{\bf S}_L \mathcal{H}_{\rm rot} \mathcal{P}_{\rm rot}
,
\label{eq:beta-stat}
\ee
implicitly fixes $\beta_{\rm rot}$ according to standard arguments of
statistical mechanics.
Upon integrating out the reservoir degrees of freedom in Eq.~(\ref{eq:rotL}),
we thus find the effective stroboscopic ensemble
\be
P_{\rm rot} = \frac{1}{Z_{\rm rot}} e^{-\beta_{\rm rot}^{\rm stat} H_{\rm rot}},
\label{eq:Protl}
\ee
where $Z_{\rm rot}$ is a normalisation and now the rotating frame Hamiltonian
is, up to $\delta J$ corrections and boundary effects,
\be
H_{\rm rot}
=
-\sum_{j=1}^{\ell - 1} \widetilde{\bf S}_j^\top \bar{J}_j^\ptop \widetilde{\bf S}_{j+1}^\ptop
+ \sum_{j=1}^\ell \left( \hat{\bf x} - \omega \hat{\bf z}\right) \cdot
\widetilde{\bf S}_j^\ptop
.
\label{eq:Hrotl}
\ee

\begin{figure*}[!ht]
  \includegraphics[width=0.95\linewidth]{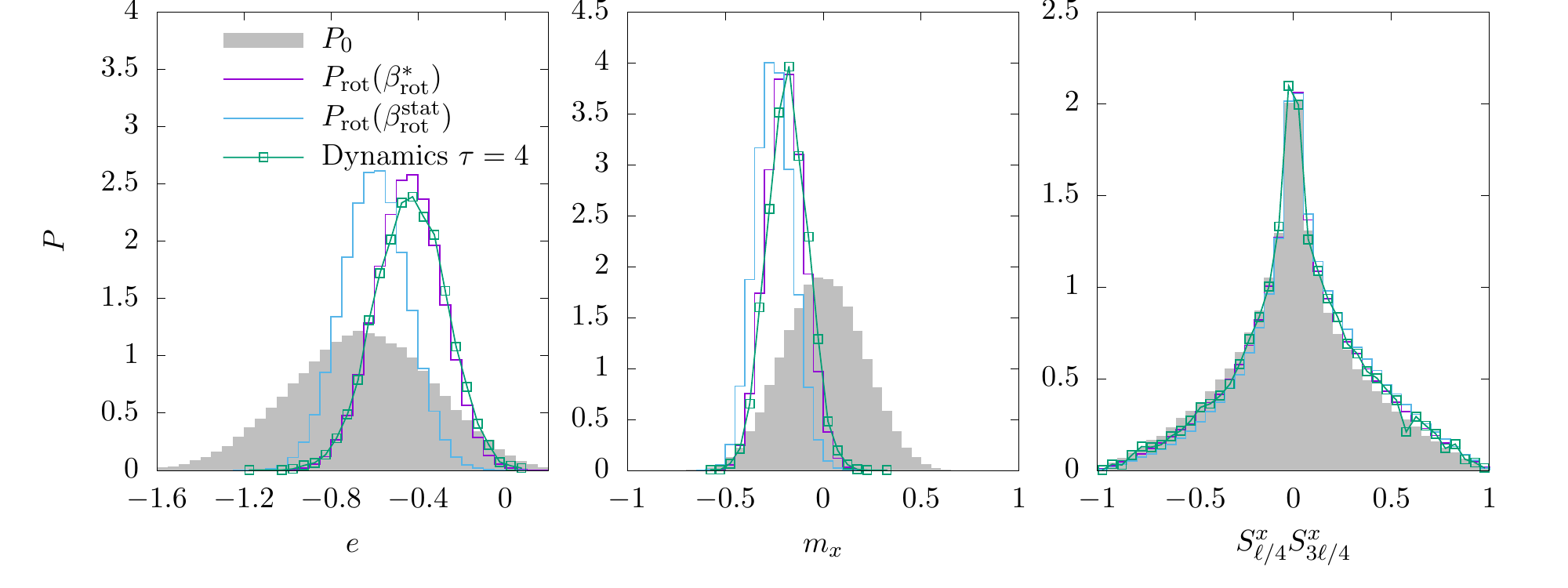}
  \caption{
    Low-frequency picture.
    Histogram distributions for observables of the system proper at
    $\tau=4$: comparing $\mathcal{N}=6000$ dynamical evolutions sampled at
    $t=1000\tau$, with $M=256,000$ MC samples from ensemble of
    Eq.~(\ref{eq:Protl}) for both $\beta_{\rm rot}=\beta_{\rm rot}^{\rm
        stat}$ and best fit $\beta_{\rm rot}^*$.
    For reference, the initial ensemble $P_0$ of Eq.~(\ref{eq:H0}) is shown in
    grey.  From left to right, the plots show energy density, $x$-magnetisation
    density, and a two-point spatial correlations function of the $x$-components
    of the spins at $j=\ell/4$ and $j=3\ell/4$.
    Here we have set $\ell=40$, $L=4000$, $\delta J = 10^{-3}$, $e_{\rm
    initial}=-0.66$.
  } \label{fig:beta-rot-comp}
\end{figure*}

As reported in Ref.~[\onlinecite{short-paper}], the ensemble of
Eq.~(\ref{eq:Protl}) yields excellent results for the distribution of
system-proper observables at very low frequencies e.g. $\tau = 10$.
For higher frequencies, however, we find in Fig.~\ref{fig:beta-rot-comp} that
these ensemble distributions deviate significantly from  the dynamical ones.
Notably, however, the ensemble of Eq.~(\ref{eq:Protl}) still reproduces the
dynamical distributions accurately if $\beta_{\rm rot}^{\rm stat}$ is replaced
with some $\beta_{\rm rot} = \beta_{\rm rot}^* < \beta_{\rm rot}^{\rm stat}$,
which can be found by fitting only the mean energy density of the system
proper.
These discrepancies highlight the limitations of the assumptions made in
writing down Eq.~(\ref{eq:beta-stat}), which we explore further in
Sec.~\ref{sec:reservoir}.

\subsubsection{Higher order corrections}

We have so far discarded corrections in $\delta J$, assuming their effects are
perturbatively small, but did not yet justify this simplification.  In the case
of high-frequency driving, the Magnus expansion provides a controlled scheme
where corrections are asymptotically small in $\tau$.
This technique can still be applied in the low-frequency regime to obtain
systematic corrections to the leading-order picture discussed above.
To this end, we observe that the full equations of motion, retaining
corrections neglected in Eq.~(\ref{eq:rotdrop}), in the rotating frame are
given by
\beA
\frac{\d \widetilde{\bf S}_j}{\d t}
&= -(\bar{J}_{j-1} \widetilde{\bf S}_{j-1} + \bar{J}_j \widetilde{\bf S}_{j+1}) \times \widetilde{\bf S}_{j} \\
&
-(\delta {J}_{j-1}(t) \widetilde{\bf S}_{j-1} + \delta{J}_j(t) \widetilde{\bf S}_{j+1}) \times \widetilde{\bf S}_{j}\\
&+ \left\{\begin{array}{lr} \hat{\bf x} - \omega \hat{\bf z}, & {\rm if}\ 1 \leq j \leq \ell, \\
  -\omega \hat{\bf z},                   & {\rm if}\ \ell < j \leq L
\end{array}
\right\}
\times
\widetilde{\bf S}_j,
\eeA
where
\beA
\bar{J}_j &=  \begin{pmatrix} \frac{1}{2}\left(J_j^x + J_j^y\right) & 0 \\ 0 & \frac{1}{2}\left(J_j^x + J_j^y\right) & 0 \\ 0 & 0 & J_j^z \end{pmatrix},
\\
\delta J_j(t) &= \frac{\Delta J_j}{2} \begin{pmatrix} -\cos (2\omega t) & \sin (2\omega t) & 0 \\ \sin (2\omega t) & \cos (2\omega t) & 0 \\ 0 & 0 & 0 \end{pmatrix}
,
\\
\Delta J_j &= J_j^y - J_j^x
.
\eeA
These equations of motion are generated by the Hamiltonian
\beA
\mathcal{H}_{\rm rot}(t) &=
-\sum_{j=1}^L
\left[
  \widetilde{\bf S}_j^\top \left(\bar{J}_j^\ptop + \delta J_j^\ptop(t) \right) \widetilde{\bf S}_{j+1}^\ptop
  +\omega\hat{\bf z} \cdot \widetilde{\bf S}_j^\ptop
  \right]
\\
& +
\sum_{j=1}^\ell
\hat{\bf x} \cdot \widetilde{\bf S}_j^\ptop
.
\label{eq:Hrott}
\eeA
We see that time-dependent corrections are formally of order $\delta J$.
Whilst we are not in the high-frequency regime, we nonetheless have an energy
scale parametrically smaller than the driving frequency $\omega \gg \delta J$.
Thus, over the course of one cycle of the drive, we can average over the
dynamics induced by the oscillating contributions proportional to $\delta J$.
That is, we may employ the Floquet-Magnus expansion in the \emph{rotating}
frame, which will generate an asymptotic series in powers of $\delta J/\omega$.

After some algebra, we find the first order global Floquet Hamiltonian in the
rotating frame,
\beA
\mathcal{H}_{F}^{\rm rot} &=
\mathcal{H}_{\rm rot}^{(0)} + \mathcal{H}_{\rm rot}^{(1)} + \mathcal{O}(\tau^2) ,\\
\mathcal{H}_{\rm rot}^{(0)} &=
- \sum_{j=1}^L \left( \widetilde{\bf S}_j^\top J_j^\ptop \widetilde{\bf S}_{j+1}^\ptop + \omega\hat{\bf z}\cdot \widetilde{\bf S}_j^\ptop \right)
+ \sum_{j=1}^\ell \hat{\bf x}\cdot \widetilde{\bf S}_j^\ptop,\\
\mathcal{H}_{\rm rot}^{(1)}
&=
- \frac{1}{16\omega}
\sum_{j=1}^L
\left\{
\Delta J_j \Delta J_j
\left( \sigma^z \widetilde{\bf S}_j \right)
\cdot \left( \sigma^x \widetilde{\bf S}_j \right)
\times \widetilde{\bf S}_{j+1}
\right.
\\
& \left.  +
\Delta J_{j-1} \Delta J_{j-1}
\left( \sigma^z \widetilde{\bf S}_j \right)
\cdot \left( \sigma^x \widetilde{\bf S}_j \right)
\times \widetilde{\bf S}_{j-1}
\right. \\
&\left.+
\Delta J_{j-1} \Delta J_j
\widetilde{\bf S}_j \cdot
\left[
  \left( \sigma^z \widetilde{\bf S}_{j+1} \right) \times \left( \sigma^x \widetilde{\bf S}_{j-1} \right)
  \right.\right.
  \\
  & \left.\left.
  -
  \left( \sigma^x \widetilde{\bf S}_{j+1} \right) \times \left( \sigma^z \widetilde{\bf S}_{j-1} \right)
  \right]
\right\},
\label{eq:HrotF}
\eeA
where we have introduced the matrices
\be
\sigma^z = \begin{pmatrix} 1 & 0 & 0 \\ 0 & -1 & 0 \\ 0 & 0 & 0 \end{pmatrix}
,\qquad
\sigma^x = \begin{pmatrix} 0 & 1 & 0 \\ 1 & 0 & 0 \\ 0 & 0 & 0 \end{pmatrix}
.
\ee
It is natural to ask whether we can distinguish these corrections coming from
$\mathcal{H}_{\rm rot}^{(1)}$ at the level of non-equilibrium ensembles.
Upon recalling that $\Delta J_j \sim \delta J$, one would expect the expansion
to have a leading contribution with a pre-factor $\delta J/\omega$, which we
would expect to be distinguishable in distributions of observables.
For the specific drive considered, however, we see from Eq.~(\ref{eq:HrotF})
that these corrections vanish and the leading terms are in fact of order
$(\delta J)^2/\omega$.
Hence, for any practically relevant frequency regime, higher-order corrections
may be safely neglected.

\section{Nature and r\^{o}le of the reservoir}\label{sec:reservoir}

Here we expand on the r\^{o}le that the reservoir plays in stabilising the
stroboscopic non-equilibrium steady states constructed in
Section~\ref{sec:driven}.
We first demonstrate how finite disorder rapidly causes runaway heating in the
bare driven system away from high frequencies.
Coupling a reservoir establishes a channel for heat transport away from the
driven sites.
Thus, at high frequencies, residual heating is compensated by dissipation with
the reservoir acting as a nearly reversible heat sink.

At low frequencies, however, the state of the reservoir is altered on a
macroscopic scale, while the spins of the system proper synchronise with the
drive.
This effect suppresses the net heat uptake of the entire spin chain and
stabilises a non-trivial steady state of the system proper at low and
intermediate frequencies.
To corroborate this picture, we evaluate spatially resolved observables and
explore the inhomogeneous distribution of magnetisation within the reservoir.
Finally, we comment on the non-Markovian nature of the reservoir in the
low-frequency regime, and implications for modelling the dynamics of the system
proper via dissipative equations of motion.

\subsection{Closed system} \label{sec:closed}

\begin{figure*}[!ht]
  \includegraphics[width=0.95\linewidth]{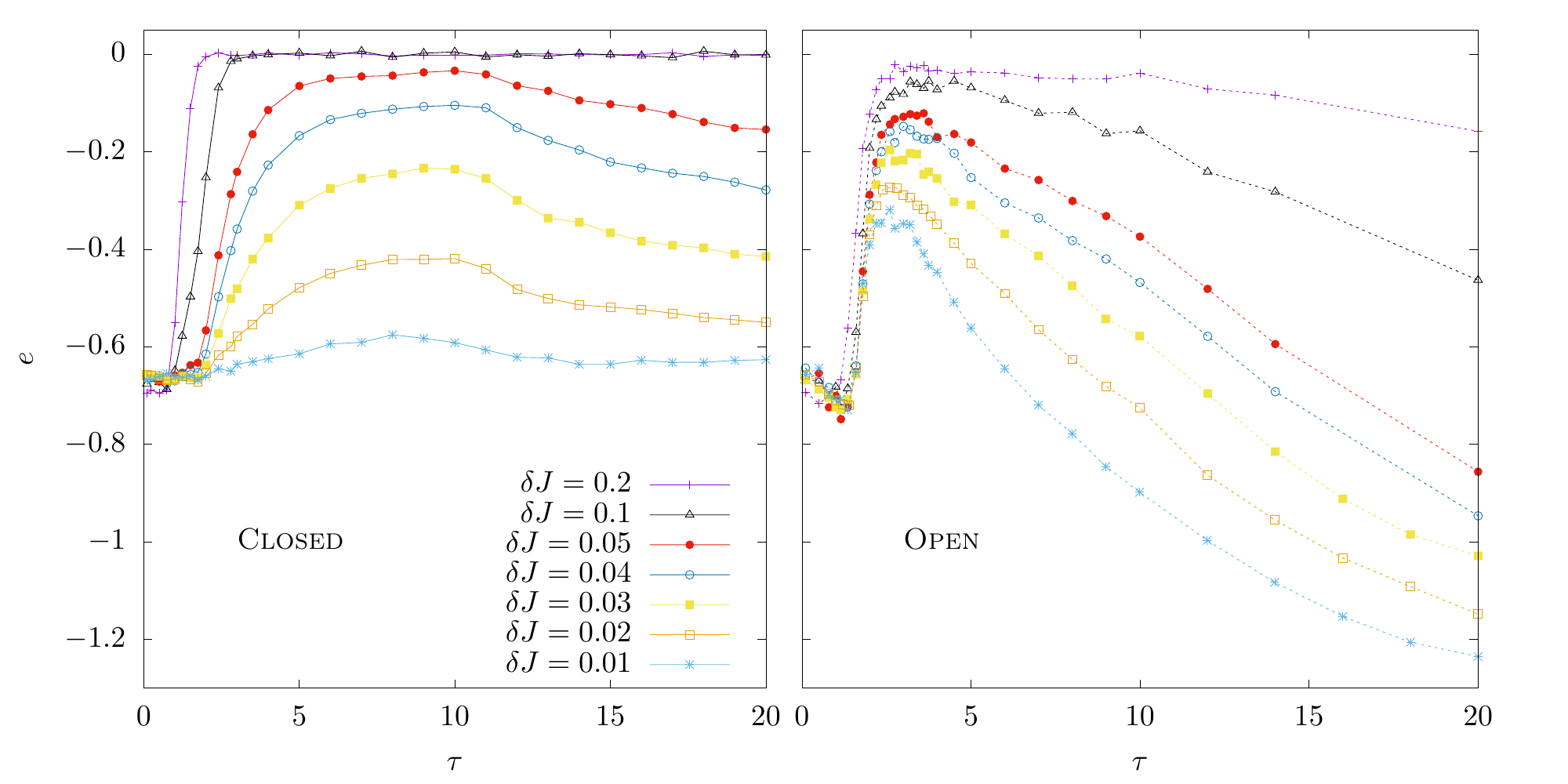}
  \caption{
    Energy density of driven sites for a closed spin chain (left, $\ell=40$) and an
    open one (right $\ell=20$, $L=2000$).
    For the dynamical distributions, due to the symmetric structure of the
    Hamiltonian, energy density $e=0$ corresponds to the `infinite temperature'
    ensemble.
    In both plots, the strength of the disorder $\delta J$ increases from bottom to
    top, which leads to an increasingly steep rise in energy absorption around
    $\tau\approx 1$.
    The initial energy density has been chosen as $e_{\rm initial}=-0.66$, and
    curves are averages over $\mathcal{N}=8000$ trajectories at $t=1000\tau$.
  } \label{fig:heating}
\end{figure*}

To understand how the reservoir modifies the dynamics, we calculate the mean
energy absorption of the driven system proper with and without a reservoir.
The results of this analysis, which are shown in Fig.~\ref{fig:heating},
suggest that the high-frequency regime is `universal' in that the leading-order
Magnus physics is insensitive to both weak disorder $\delta J$ and the presence
of a reservoir.
This behaviour can be intuitively understood from the structure of the
Floquet-Magnus expansion discussed in Sec.~\ref{sec:driven}.
To this end, we may divide the Hamiltonian of the entire spin chain into a
system-proper and a reservoir contribution
\beA
\mathcal{H}(t) &= H^S(t) + H^R, \\
H^S(t) &= -\sum_{j=1}^{\ell-1} {\bf S}_j^\top J_j^\ptop {\bf S}_{j+1}^\ptop
+ \sum_{j=1}^\ell {\bf B}(t) \cdot {\bf S}_j^\ptop, \\
H^R &= -\sum_{j=\ell}^{L} {\bf S}_j^\top J_j^\ptop {\bf S}_{j+1}^\ptop .
\eeA
The two lowest-order terms of the Floquet-Magnus expansion are then given by
\beA
H_F^{(0)} &= \frac{1}{\tau} \int_0^\tau \d t\, \mathcal{H}(t) = \mathcal{H}_0,\\
H_F^{(1)} &= h^{(1)}_{F,\ {\rm loc}} + h^{(1)}_{F,\ {\rm int}}, \\
h^{(1)}_{F,\ {\rm loc}} &= \frac{1}{2!\tau} \int_0^\tau \d t_1 \int_0^{t_1} \d t_2\, \left\{ H^S(t_1), H^S(t_2) \right\} \\
h^{(1)}_{F,\ {\rm int}} &= \frac{1}{2!\tau} \int_0^\tau \d t_1 \int_0^{t_1} \d t_2 \\
&\left(
\left\{ H^R, H^S(t_2) \right\}
+
\left\{ H^S(t_1), H^R \right\}
\right).
\eeA
Here, $h^{(1)}_{F,\ {\rm loc}}$ depends solely on degrees of freedom of the
system proper.
Furthermore, since only nearest-neighbour spins interact, $h^{(1)}_{F,\ {\rm
int}}$ depends only on the spins adjacent to the boundary of the system proper.
By extension, since the $N$th-order correction $H_F^{(N)}$ involves $N$ nested
Poisson brackets, any modification of the reservoir Hamiltonian a distance $M$
away from the system proper is suppressed by a factor of $\tau^M$.
As a result, the reservoir is only significantly affected by the driving in
close vicinity to the system proper at sufficiently high frequencies.
The small tails visible in Fig.~\ref{fig:sketch} arise from such corrections.
Consequently, we can expect net energy absorption to be small and a local
picture to be sufficient for the description of the system proper in the
high-frequency regime.

In the low-frequency regime, the Floquet-Magnus expansion is applicable only in
the rotating frame, where a virtual magnetic field proportional to $\omega$
acts on the entire reservoir, see Sec.~\ref{sec:low-freq}.
Therefore, the driving eventually affects spins arbitrarily far from the system
proper, thus inducing a large-scale redistribution of energy, which explains
the qualitative difference between the behaviour of the closed and the open
system seen at low frequencies in Fig.~\ref{fig:heating}.
While the closed system rapidly approaches a trivial infinite-temperature state
with increasing strength of the disorder $\delta J$, long-range energy transfer
to the reservoir enables the stabilisation of a synchronised steady state,
whose mean energy may even fall below its initial value in the open system.
The lifetime of this steady state can be expected to scale with the size of the
reservoir as it is limited by the heat capacity of the entire spin chain rather
than the system proper.

\subsection{Synchronisation} \label{sec:sync}
\begin{figure*}[!ht]
  \includegraphics[width=\linewidth]{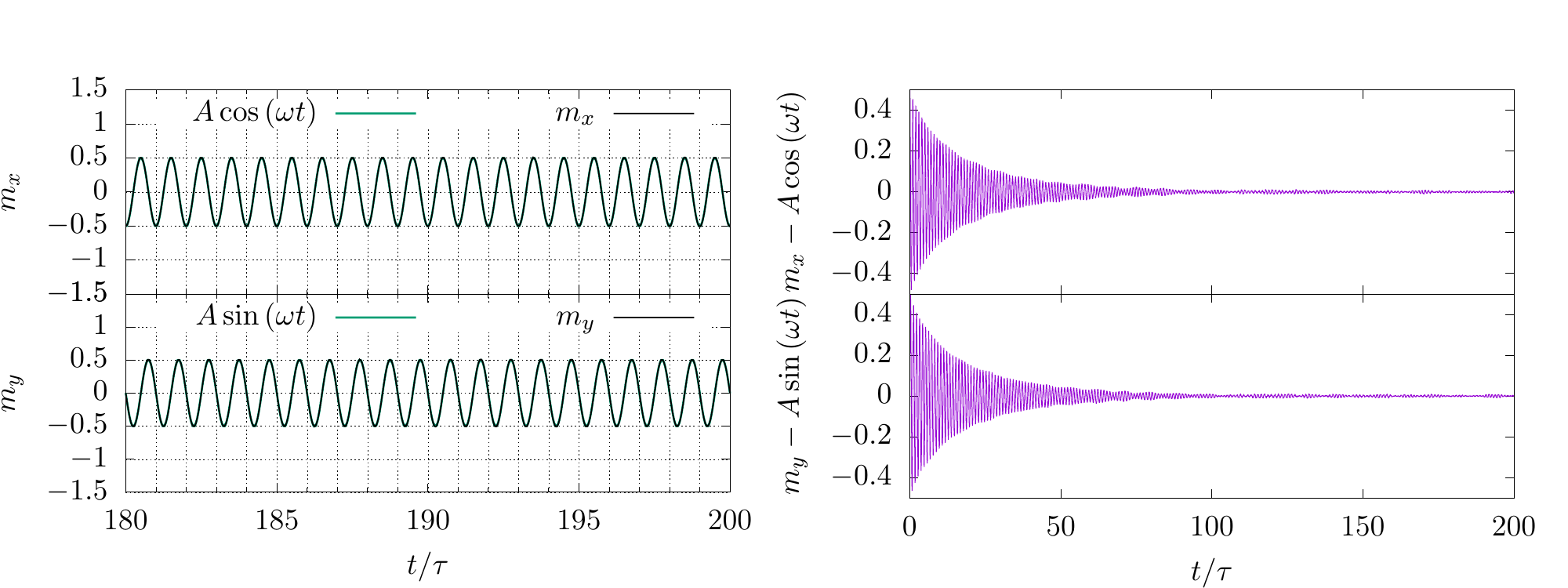}
  \caption{
  Synchronisation of open system.
  Left: local magnetisations $m_x$ and $m_y$, compared with the scaled
  magnetic field projected onto the corresponding direction, where the scaling
  factor $A=0.501$ has been fitted.
  Right: difference between these functions. There is a transient period of
  approximately 100 cycles, after which the system is well synchronised up to
  small errors.
  Here, we have set $\tau = 10$, $\ell = 40$, $L=2000$, $e_{\rm
        initial}=-0.66$, $\delta J =10^{-3}$ and simulations are averaged over
  $\mathcal{N}=800$ trajectories.
  } \label{fig:closed-sync}
\end{figure*}

\begin{figure*}[!ht]
  \includegraphics[width=\linewidth]{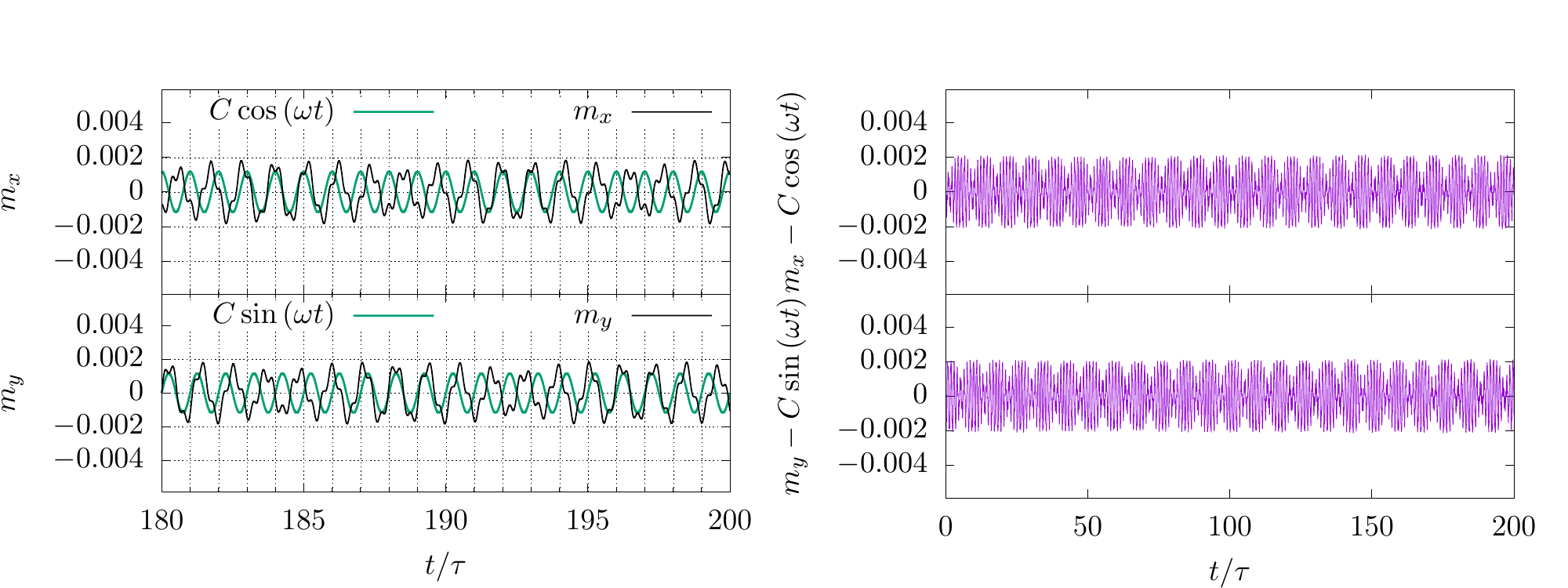}
  \caption{
    Lack of synchronisation of closed system.
    Left: local magnetisations $m_x$ and $m_y$, compared with the scaled
    magnetic field projected onto the appropriate direction, where the scaling
    factor $C=-0.00117$ has been fitted.
    Right: difference between these functions.
    No synchronisation is observable over any significant timescale.
    Here, we have set $\tau = 10$, $\ell=L=40$, $e_{\rm initial} = -0.66$,
    $\delta J = 10^{-3}$, and simulations are averaged over $\mathcal{N}=4000$
    trajectories.
  } \label{fig:open-sync}
\end{figure*}

The relationship between heating and synchronisation can be described
intuitively by examining the continuous-time dynamics of the driven spin chain.
To this end, we first note that the total energy absorption over the time $t$
can be expressed as
\beA
L\left[\mathfrak{e}(t) - \mathfrak{e}_{\rm initial}\right]
&=
\int_0^t \d s\left\langle \frac{\d}{\d s} \mathcal{H}(s) \right\rangle_0 \\
&=
\ell \int_0^t \d s\, \dot{\bf B}(s) \cdot \langle {\bf m} \rangle_0 ,
\label{eq:sync-condition}
\eeA
where $\langle\cdots\rangle_0$ denotes the average over initial states, the
magnetisation ${\bf m}$ of the system proper is defined according to
Eq.~(\ref{eq:el}), and $\mathfrak{e}$ is the energy density of the entire
system.
Hence, the total energy absorption vanishes if $\langle {\bf m}(s) \rangle_0
  \propto {\bf B}(s)$ for all times.
Fig.~\ref{fig:open-sync} shows that, after a transient phase of about 100-200
cycles, the open system, indeed enters a state where this synchronisation
condition is nearly met and the rate of energy absorption is minimal.
We note that the remaining deviations, which cause residual heating in the
slow-driving regimes, cannot be explained by a constant phase lag.

By contrast, no synchronisation takes place in the closed system as shown in
Fig.~\ref{fig:closed-sync}; instead the deviations from the synchronisation
condition are homogeneous on a scale $t\sim 11\tau$.
This timescale comes from the beating frequency $\sqrt{1+\omega^2}-\omega$ of
the exact solution for the closed system magnetisation in the clean limit
$\delta J = 0$.
At the same time, these deviations are of the same order of magnitude as in the
synchronised state of the open system, containing a statistical contribution
from finite sampling of the initial conditions as well as deviations induced by
finite $\delta J$.

To understand these observations, we may again invoke the rotating-frame picture
of Sec.~\ref{sec:low-freq}, where the Hamiltonian of the entire system is given
by Eq.~(\ref{eq:Hrott}) and the system proper is subject to the effective magnetic
field ${\bf B}_{\rm eff} = \hat{\bf x} - \omega\hat{\bf z}$.
The magnetisation of the system proper parallel to ${\bf B}_{\rm eff}$ is
conserved up to corrections of order $\delta J$, and therefore does not change
significantly on relevant time scales.
This almost-conservation law is broken once the system proper is coupled to
the reservoir.
As a result, a persistent magnetisation builds up in the $x$-direction as the
entire system relaxes to the effective Gibbs ensemble (\ref{eq:rotL}).
In the lab frame, this process corresponds to the synchronisation observed in
Fig.~\ref{fig:open-sync}.
It thus also becomes clear that the deviations from the synchronisation
condition that persist in both the closed and the synchronised open system
should be attributed to the disorder in the coupling constants, which renders the
rotating-frame Hamiltonian a non-conserved quantity, regardless of the presence
of the reservoir.
Finally, since the open system is able to redistribute energy into the
reservoir, its effective heating rate, incurred by the residual energy
absorption due to imperfect synchronisation, is strongly suppressed compared to
the closed system.

\subsection{Profiles} \label{sec:prof}

\begin{figure*}[!ht]
  \includegraphics[width=0.45\linewidth]{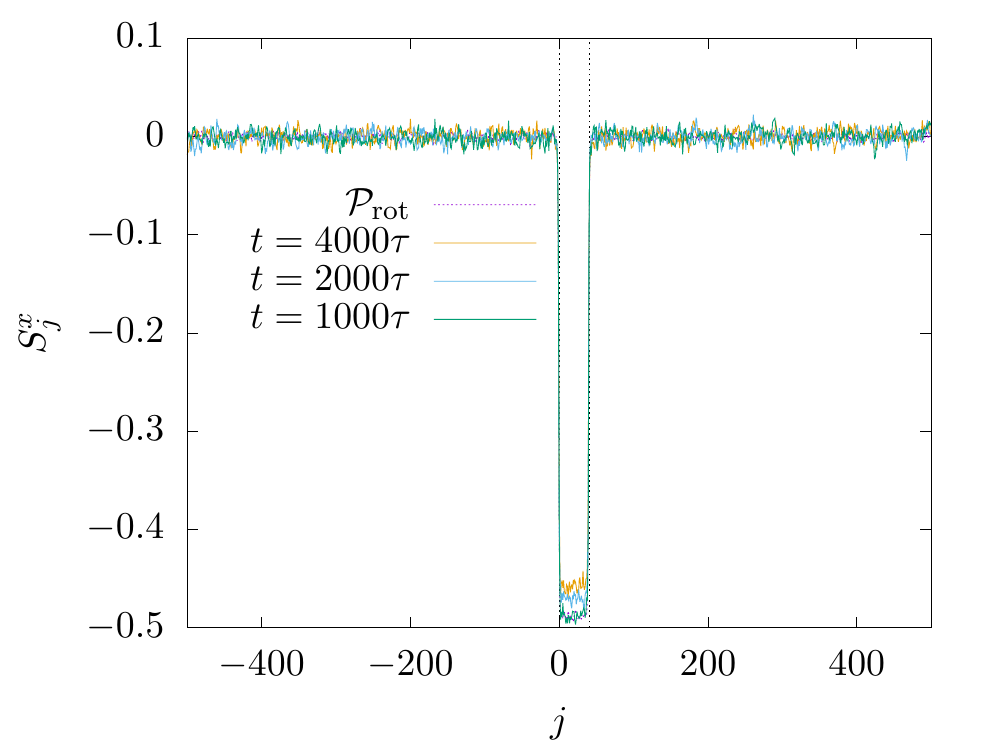}
  \includegraphics[width=0.45\linewidth]{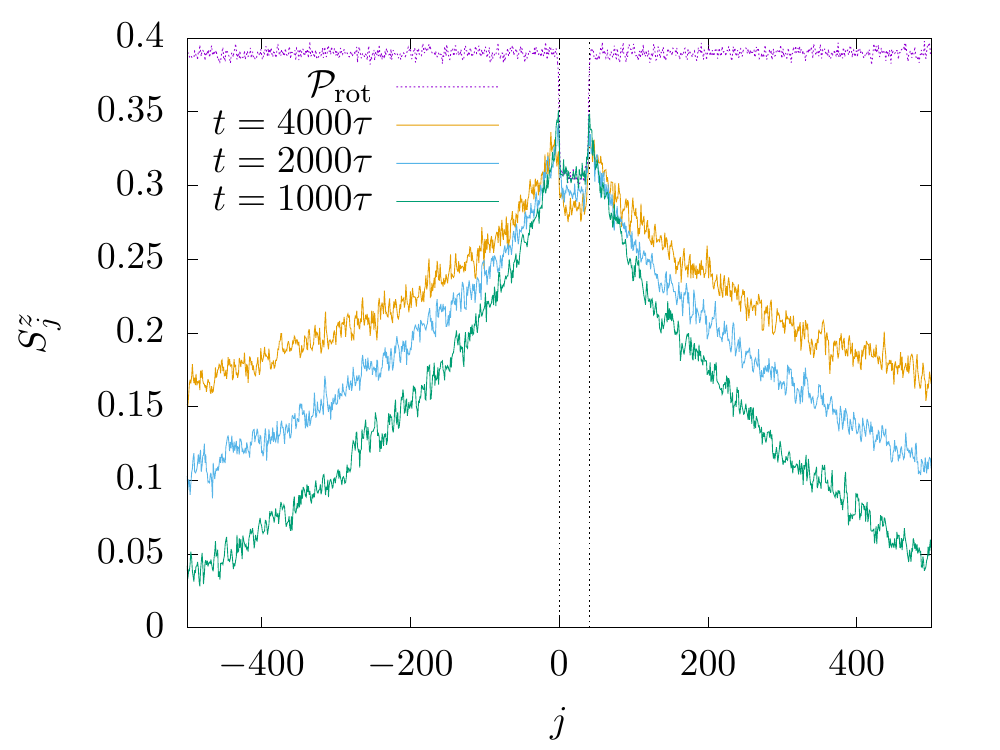}
  \caption{
    Spatially resolved profiles for $S_j^x$ and $S_j^z$ in the slow-driving
    regime $\tau=10$ for $\ell=40$, $L=2000$, $\delta J = 10^{-3}$, $e_{\rm
    initial}=-0.66$.
    Shown are the dynamical results, averaged over $\mathcal{N}=8000$
    trajectories, for various $t$, compared with the rotating-frame ensemble
    $\mathcal{P}_\mathrm{rot}$ of Eq.~(\ref{eq:rotL}) averaged over
    $\mathcal{M}=32000$ MC samples.
    Boundaries of the system proper are indicated by vertical dashed lines.
  } \label{fig:prof}
\end{figure*}

\label{sec:drift}
Our understanding of the low-frequency regime rests on the assumption that in
the rotating frame the system relaxes to the global Gibbs state of
Eq.~(\ref{eq:rotL}) on a timescale that is well separated from the heating
timescale determined by $\delta J$.  Given the form of the $z$-field present in
the Hamiltonian of Eq.~(\ref{eq:Hrot-L}), one might expect that this relaxation
happens uniformly throughout the reservoir. 
As a matter of causality, however, the reservoir spins far away from the driven
sites cannot be instantly affected by the driving.
In fact, the interaction strength sets the scale for a finite group velocity
$v\sim J=1$, and any correlation function involving solely  degrees of freedom
separated by a distance $r$ cannot distinguish between the driven and undriven
systems outside of the light-cone $r>vt$.
Hence, the rearrangement of local degrees of freedom and the large-scale spread
of correlations must occur on different timescales.

To uncover the inhomogeneous nature of the relaxation process explicitly, we
calculate the spatial profiles of the local observables $S_j^x$ and $S_j^z$.
These data, plotted in Figs.~\ref{fig:sketch} and \ref{fig:prof}, show that for
low-frequency driving the $x$- and $z$-magnetisations of the system proper have
settled to the flat profiles predicted by the local Gibbs state of
Eq.~(\ref{eq:Protl}) after around 1000 cycles.
At this time, the $z$-magnetisation profile of the reservoir is neither
homogeneous nor stationary.

To understand the mechanism of the subsequent global relaxation process, we may
observe that the $z$-magnetisation is a locally conserved quantity everywhere
in the bulk of the reservoir, up to corrections of order $\delta J$.
The breaking of this conservation law at the boundary of the reservoir, where
the system proper generates an oscillating magnetic field perpendicular to
$\hat{\bf z}$, leads to gradual build-up of the $z$-magnetisation profile.
This profile spreads, presumably diffusively, into the reservoir and eventually
becomes flat when the system has fully relaxed to the global Gibbs state of
Eq.~(\ref{eq:rotL}).
At the same time, deviations from synchronisation in the system proper cause
small heat currents to flow into the reservoir and elevate its overall energy
density on the heating timescale set by $\delta J$.

We may now return to the question of how to determine the effective inverse
temperature $\beta_{\rm rot}$ for the rotating-frame ensemble encountered in
Sec.~\ref{sec:low-freq}.
According to statistical mechanics, this parameter should be fixed by energy
conservation, as indicated by Eq.~(\ref{eq:beta-stat}).
However, this approach assumes that the system has fully relaxed to the global
Gibbs state (\ref{eq:rotL}), which, as shown by Figs.~\ref{fig:sketch} and
\ref{fig:prof}, is clearly not the case for the evolution times considered in
Fig.~\ref{fig:beta-rot-comp}.
It is therefore surprising that $\beta_{\rm rot}$ as determined by
Eq.~(\ref{eq:beta-stat}) yields excellent agreement between ensemble and
dynamical distributions at very low frequencies \cite{short-paper}.
A posteriori, this result may be attributed to the fact that the virtual
magnetic field on the reservoir, and thus the amplitude of the emerging
magnetisation profile are small for $\omega\ll 1$.
Thus the initial state of the reservoir does not deviate substantially from its
rotating-frame Gibbs state. Away from ultra-low frequencies, agreement between
ensemble and dynamical distributions may still be achieved by fitting
$\beta_{\rm rot}$ to the mean energy of the system proper, as we have shown in
Sec.~\ref{sec:low-freq}.
This observation suggests that the reservoir locally reaches an effective
equilibrium state at its boundary with the system proper on a much smaller
timescale than that of the global relaxation process.

\begin{figure*}[!ht]
  \includegraphics[width=0.45\linewidth]{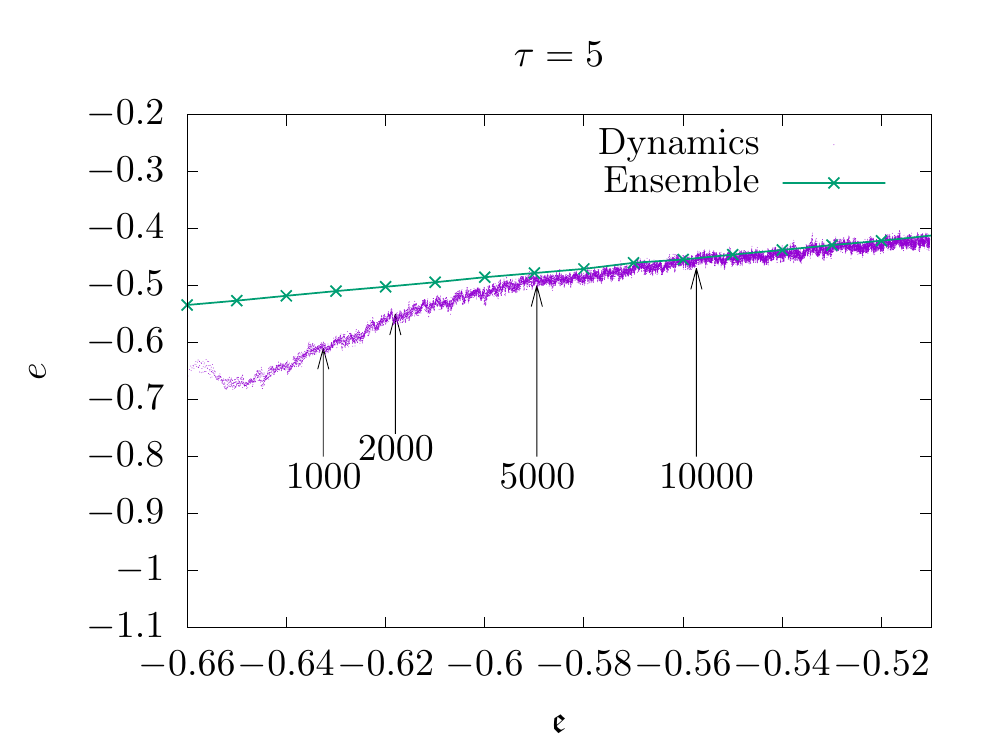}
  \includegraphics[width=0.45\linewidth]{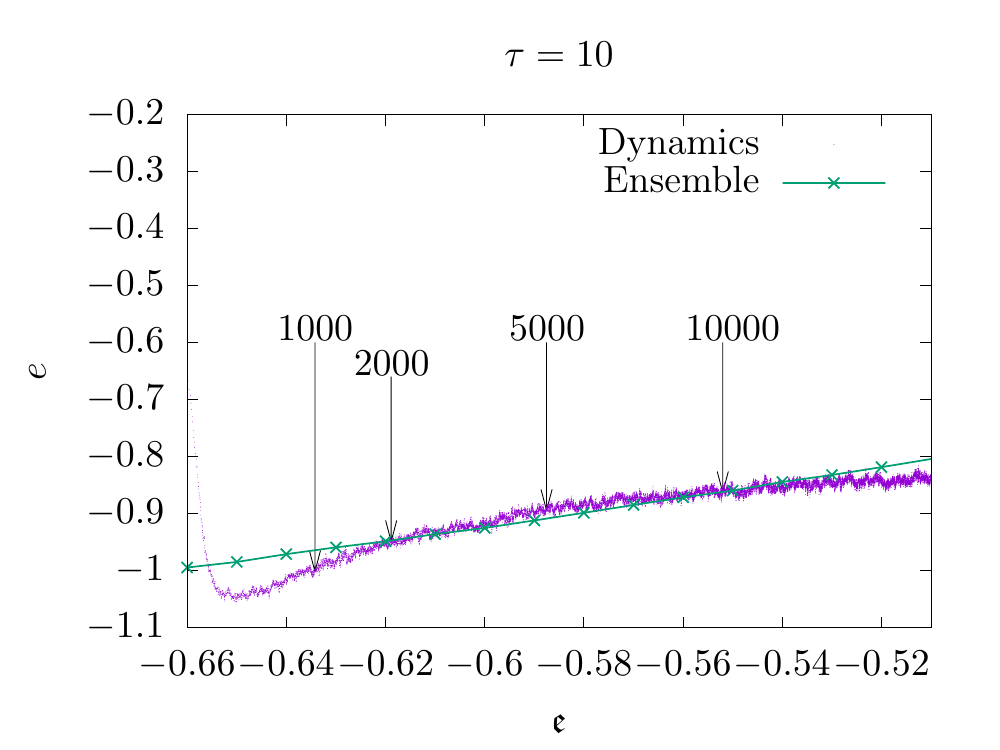}
  \caption{
    Implicit relationship between the energy density of the
    total system $\mathfrak{e}$ and the system proper only $e$.
    The green curve is produced by sweeping $\beta_{\rm rot}$ and
    evaluating the energy density of both system proper and total system
    from the ensemble of Eq.~(\ref{eq:rotL}).
    The purple data is an average of $\mathcal{N} = 8000$ dynamical evolutions
    with $\ell=40$, $L=4000$, $\delta J = 10^{-3}$, $e_{\rm initial}=-0.66$.
    The number of cycles elapsed are indicated by arrows.
  } \label{fig:drift}
\end{figure*}
This reasoning implies an implicit relationship between the energy density of
the system proper $e$, and that of the entire system $\mathfrak{e}$, which can
be probed quantitatively.
In the rotating-frame ensemble (\ref{eq:rotL}), both $e$ and $\mathfrak{e}$ are
determined by a single parameter, $\beta_{\rm rot}$.
If we assume that relaxation to this ensemble occurs on a much faster timescale
than heating, we may described the system by an effective rotating-frame Gibbs
state with inverse temperature $\beta_{\rm rot} = \beta_{\rm rot}(t)$ after
some transient period.
We may thus sweep $\beta_{\rm rot}$ through a range of values and plot the
energy densities $\mathfrak{e}(\beta_{\rm rot})$ against $e(\beta_{\rm rot})$.
This plot can then be compared with the actual values of these quantities
obtained from dynamical simulations.
The results of this analysis are shown in Fig.~\ref{fig:drift}.  For $\tau=10$,
the dynamical energy densities come fairly close to values predicted by the
effective ensemble for simulation times $t\gtrsim 1000\tau$.
In this regime, $\beta_{\rm rot}$ can be accurately determined from energy
conservation, which yields excellent results as exhibited in
Ref.~[\onlinecite{short-paper}].
For $\tau=5$, Fig.~\ref{fig:drift} shows that this approach is viable only for
$t\gtrsim 5000\tau$, which explains why $\beta_{\rm rot}$ required fitting in
order to achieve agreement between dynamical and ensemble distributions as seen
in Fig.~\ref{fig:beta-rot-comp} where $t=1000\tau$.

\subsection{Non-Markovianity}
Throughout this paper, we have described the reservoir by explicitly simulating
its full microscopic dynamics.
We have, however, not yet considered the option of accounting for the reservoir
by modifying only the equations of of the system proper, which might
drastically reduce the computational cost of numerical simulations.
A full analysis of this approach would go beyond the scope of this work.
On the basis of our results we can, however, identify some of its immediate
limitations.

The simplest way to construct the dissipative dynamics of the system proper
would, arguably, be to replace the equations of motion for the edge spins with
the Langevin equation \cite{Ma-Dudarev-Semenov-Woo,Brown}
\beA
\frac{\d {\bf S}_j}{\d t}
& =
-{\boldsymbol\Omega}_j  \times {\bf S}_j
-\gamma_s {\bf S}_j \times \left(  {\bf S}_j \times{\boldsymbol\Omega}_j  \right) \\
&+ {\bf h}_j(t)  \times {\bf S}_j
,\qquad j=1,\ell ,
\label{eq:langevin}
\eeA
and for the remaining sites $j=2,\ldots,\ell-1$ the equations of motion
Eq.~(\ref{eq:eom}) remain unchanged.
Here,
${\bf h}_j(t)$ represents a delta-correlated noise vector 
$\langle h_{i,\alpha}(t) h_{j,\beta}(t') \rangle =
\mu_s\delta_{ij}\delta_{\alpha\beta} \delta(t-t')$,
$\mu_s$ is the noise strength, and $\gamma_s$ is a damping constant.
To mimic a thermal reservoir at inverse temperature $\beta$, these quantities
must obey the fluctuation-dissipation relation $2\gamma_s = \mu_s \beta$.
If the effective magnetic field ${\boldsymbol\Omega}_j$ is time independent,
the system proper then relaxes to the Gibbs state (\ref{eq:H0}).

One might now include the driving by adding the on-sit magnetic field 
${\bf B}(t)$ to the effective field ${\boldsymbol\Omega}_j$.
However, our results show that this modification would invalidate the basic
assumptions underpinning the Langevin equation (\ref{eq:langevin}).
Specifically, the Markovian limit i.e. delta-correlated noise, is realised only
if the reservoir constantly remains in its initial equilibrium state and its
correlation functions decay fast on the observational timescale.
As seen in Figs.~\ref{fig:sketch} and \ref{fig:prof}, however, the full
dynamical simulation suggests that, even for $L\to\infty$, an ever-growing
region of the reservoir departs from its initial state while correlations with
the system proper build up continuously.
One should therefore not expect a simple Langevin model with additively
incorporated driving to reproduce the dynamics of the system proper accurately,
as least beyond the limit of ultra-low frequencies, which is characterised by
relaxation to an instantaneous Gibbs state.

In fact, our results show that any attempt to describe the system-proper
dynamics at intermediate frequencies by means of dissipative equations of
motion would have to give up on the assumption of Markovianity.
Whether or not it is still possible to derive accurate and tractable
non-Markovian time-evolution equations, e.g. by using Nakajima-Zwanzig
projection-operator techniques, remains as an important subject for future
research \cite{Nakajima,Zwanzig,Breuer}.

\section{Perspectives}
The central aim of this paper was to shed new light on the physics of classical
many-body systems that are subject to periodic driving while being coupled to a
large thermal reservoir. 
To make progress in this direction, we have modelled both the driven system and
the reservoir as spin chains with nearest-neighbour interactions and weak
disorder. 
Since the full Hamiltonian dynamics of this setup can be simulated exactly at
moderate numerical cost, we were able to avoid the use of dissipative equations
of motion that account for the reservoir in a phenomenological or approximate
way. 
While some of our more quantitative results may be contingent on the specific
setting we have considered, we expect our main insights to be representative for
a broader class of systems.

In particular, a high-frequency regime, where energy absorption is strongly
suppressed and the stroboscopic dynamics of the driven system is governed by its
averaged Hamiltonian, plus leading corrections obtained from the
classical Floquet-Magnus expansion, should generically exist. 
Our results corroborate the natural expectation that, in this regime, the
reservoir acts, up to small perturbations at its boundary with the driven
system, as a nearly reversible heat sink balancing residual energy uptake from
the drive. 
Perhaps more surprisingly, we found that this behaviour changes quite abruptly as
the driving frequency decreases below some threshold value, which is determined
by the typical energy scale of the system and, to a lesser degree, by the
strength of the disorder, see Fig.~\ref{fig:heating}. 
The system then enters a crossover regime, which is characterised by a sharp
increase in energy absorption and covers only a small range of frequencies. 
Understanding the microscopic mechanism of this crossover as well as its
putative dependence on the dimensionality of the system, the nature of the drive
and the range of interactions, provides an intriguing and presumably challenging
subject for future research.

Upon further reducing the driving frequency, the system eventually enters a
low-frequency regime, which connects smoothly to the, most likely universal,
quasi-static limit, where the driven degrees of freedom are constantly described
by an instantaneous equilibrium state. 
This regime is characterised by rapid synchronisation between the driven system
and the applied field and, away from the quasi-static limit, a gradual
rearrangement of reservoir degrees of freedom over long distances, which is
accompanied by the steady build-up of long-range correlations. 
Here, we were able to show that this behaviour may in fact extend over a large
range of frequencies, which is limited from above only by the crossover to the
high-frequency regime. 
This analysis, however, crucially relies on the existence of a rotating
reference frame, where the Hamiltonian of the entire system is nearly time
independent, thus providing a stroboscopically conserved quantity. 
Whether or not almost conserved quantities, which prevent the system from
approaching an infinite-temperature state on a practically long time scale exist
for more general systems and driving protocols, remains an open question. 
It is, however, plausible to expect that such quantities may be constructed at
least perturbatively from the quasi-static limit, where the instantaneous
Hamiltonian serves as an adiabatic invariant. 
It would then be interesting to explore whether a qualitatively new type of
behaviour emerges between the low-frequency regime and the crossover to the
high-frequency regime and how it can be characterised. 

Finally, our work opens an interesting perspective in the area of open dynamical
systems. 
As we have briefly discussed at the end of Sec.~\ref{sec:reservoir}, the
conventional Langevin-approach to classical open-system dynamics is limited by
the assumption of a nearly invariant reservoir with fast decaying correlations
on the observational time scales. 
However, for the system we have analysed here, this condition can be met only
near the quasi-static limit, and perhaps in the high-frequency regime upon
replacing the system Hamiltonian with a suitably truncated Floquet-Magnus
Hamiltonian; in the latter case, it may be possible to develop a classical
framework similar to the stochastic-wave-function method in Floquet
representation, which provides a dynamical description for open quantum systems
subject to rapidly oscillating driving fields 
\cite{Breuer-Petruccione}. 
For intermediate frequencies, however, our results strongly suggest that
non-Markovian equations of motion will have to be adopted to describe the
dynamics of periodically driven open systems, presumably in both the classical
and the quantum case. 
Finding systematic ways to derive and analyse these equations will require
further research. Our present work provides both a starting point for such
investigations and a valuable benchmark for their results. 

\subsection*{Data access statement}
The source code used for all simulations, and all data used in figures, is
freely available at \url{https://github.com/tveness/spinchain-papers}.

\section*{Acknowledgements}
We thank Anatoli Polkovnikov for suggesting the research problem and for
thoughtful discussions.
We acknowledge support from the University of Nottingham through a Nottingham
Research Fellowship and from UK Research and Innovation through a Future Leaders
Fellowship (Grant Reference: MR/S034714/1).
TV is grateful for hospitality at Boston University during the final stages of
research.


\begin{references}
  \bibitem{Floquet-1883}
  G. Floquet,
  \href{https://doi.org/10.24033/asens.220}{Scientific annals of the École Normale
    Supérieure 12, 47 (1883)}.
  \bibitem{Kitagawa-Oka-Brataas-Fu-Demler}
  T.~Kitagawa, T.~Oka, A.~Brataas, L.~Fu, E.~Demler,
  \href{https://doi.org/10.1103/PhysRevB.84.235108}{Phys. Rev. B {\bf 84}, 235108 (2011)}.
  \bibitem{Sambe}
  H.~Sambe,
  \href{https://doi.org/10.1103/PhysRevA.7.2203}{Phys. Rev. A {\bf 7}, 2203 (1973)}.
  \bibitem{Torres-Kunold}
  M.~Torres and A.~Kunold,
  \href{https://doi.org/10.1103/PhysRevB.71.115313}{Phys. Rev. B {\bf 71}, 115313 (2005)}.
  \bibitem{Bukov-DAlessio-Polkovnikov-review}
  M.~Bukov, L.~D'Alessio, A.~Polkovnikov,
  \href{https://doi.org/10.1080/00018732.2015.1055918}{Adv. Phys. {\bf 64}, 139 (2015)}.
  \bibitem{Eckardt-rmp}
  A.~Eckardt,
  \href{https://doi.org/10.1103/RevModPhys.89.011004}{Rev. Mod. Phys. {\bf 89}, 011004 (2017)}.
  \bibitem{Anatoli-acknowledgement}
  We thank Anatoli Polkovnikov for bringing this argument to our attention.
  \bibitem{Arnold}
  V.~I~Arnold, A.~Avez, \emph{Ergodic Problems of Classical Mechanics} (1968).
  \bibitem{Broer-Krauskopf}
  H.~W.~Broer, B.~Krauskopf,
  \href{https://doi.org/10.1063/1.1337757}{AIP Conference Proceedings {\bf 548}, 31 (2000)}.
  \bibitem{Landau-Lifshitz}
  L.~D.~Landau, E.~M.~Lifshitz, \emph{Mechanics. Vol. 1} \S 30, Pergamon Press (1960)
  \bibitem{KWW}
  T.~Kinoshita, T.~Wenger, D.~S. Weiss,
  \href{https://doi.org/10.1038/nature04693}{Nature {\bf 440}, 900 (2006)}.
  \bibitem{Khemani-PRL}
  V.~Khemani, A.~Lazarides, R.~Moessner, S.~L.~Sondhi,
  \href{https://doi.org/10.1103/PhysRevLett.116.250401}{Phys. Rev. Lett. {\bf 116}, 250401 (2016)}.
  \bibitem{Curt-phase}
  C.~W.~von~Keyserlingk, S.~L.~Sondhi,
  \href{https://doi.org/10.1103/PhysRevB.93.245145}{Phys. Rev. B {\bf 93}, 245145 (2016)}.
  \bibitem{Chalker-PRX}
  A.~Chan, A.~De Luca, J.~T.~Chalker,
  \href{https://doi.org/10.1103/PhysRevX.8.041019}{Phys. Rev. X {\bf 8}, 041019 (2018)}.
  \bibitem{Weitenberg-Simonet}
  C.~Weitenberg, J.~Simonet,
  \href{https://doi.org/10.1038/s41567-021-01316-x}{Nat. Phys. {\bf 17}, 1342 (2021)}.
  \bibitem{Rudner}
  M.~S.~Rudner, N.~H.~Lindner,
  \href{https://doi.org/10.1038/s42254-020-0170-z}{Nat. Rev. Phys. {\bf 2}, 229 (2020)}.
  \bibitem{Ponte-Chandran-Papic-Abanin}
  P.~Ponte, A.~Chandran, Z.~Papic, D.~A.~Abanin,
  \href{https://doi.org/10.1016/j.aop.2014.11.008}{Ann. Phys. {\bf 353}, 196 (2015)}.
  \bibitem{Lazarides-heating}
  A.~Lazarides, A.~Das, R.~Moessner,
  \href{https://doi.org/10.1103/PhysRevE.90.012110}{Phys. Rev. E {\bf 90}, 012110 (2014)}.
  \bibitem{Russomanno-Silva-Santoro}
  A.~Russomanno, A.~Silva, G.~E.~Santoro, J. Stat. Mech. {\bf 2013}(09), P09012 (2013).
  \bibitem{Ikeda-Polkovnikov}
  T.~N.~Ikeda, A.~Polkovnikov,
  \href{https://doi.org/10.1103/PhysRevB.104.134308}{Phys. Rev. B {\bf 104}, 134308 (2021)}.
  \bibitem{Haldar-Das}
  A.~Haldar, A.~Das, 
  \href{https://doi.org/10.1088/1361-648X/ac03d2}{J. Phys.: Condens. Matter {\bf 34} 234001 (2022)}.
  \bibitem{Lazarides-Das-Moessner-periodic}
  A.~Lazarides, A.~Das, and R.~Moessner,
  \href{https://doi.org/10.1103/PhysRevLett.112.150401}{Phys. Rev. Lett. {\bf 112}, 150401 (2014)}.
  \bibitem{Abanin-DeRoeck-Ho-Huveneers}
  D.~Abanin, W.~De~Roeck, W.~W.~Ho, F.~Huveneers,
  \href{https://doi.org/10.1007/s00220-017-2930-x}{Commun.  Math. Phys. {\bf 354}, 809 (2017)}.
  \bibitem{Abanin-DeRoeck-Ho-Huveneers2}
  D.~A.~Abanin, W.~De~Roeck, W.~W.~Ho, F.~Huveneers,
  \href{https://doi.org/10.1103/PhysRevB.95.014112}{Phys. Rev. B {\bf 95}, 014112 (2017)}.
  \bibitem{Abanin-DeRoeck-Huveneers}
  D.~A.~Abanin, W.~De~Roeck, F.~Huveneers,
  \href{https://doi.org/10.1103/PhysRevLett.115.256803}{Phys. Rev. Lett. {\bf 115}, 256803 (2015)}.
  \bibitem{Mori-Kuwahara-Saito}
  T.~Mori, T.~Kuwahara, K.~Saito,
  \href{https://doi.org/10.1103/PhysRevLett.116.120401}{Phys. Rev. Lett. {\bf 116}, 120401 (2016)}.
  \bibitem{Mori-Kuwahara-Saito2}
  T.~Kuwahara, T.~Mori, K.~Saito,
  \href{https://doi.org/10.1016/j.aop.2016.01.012}{Ann. Phys. {\bf 367}, 96 (2016)}.
  \bibitem{Hodson-Jarzynski}
  W.~Hodson, C.~Jarzynski,
  \href{https://doi.org/10.1103/PhysRevResearch.3.013219}{Phys. Rev. Research {\bf 3}, 013219 (2021)}.
  \bibitem{Howell}
  O.~Howell, P.~Weinberg, D.~Sels, A.~Polkovnikov, M.~Bukov,
  \href{https://doi.org/10.1103/PhysRevLett.122.010602}{Phys. Rev. Lett. {\bf 122}, 010602 (2019)}.
  \bibitem{Nunnenkamp}
  A.~Pizzi, A.~Nunnenkamp, J.~Knolle,
  \href{https://doi.org/10.1103/PhysRevLett.127.140602}{Phys. Rev. Lett. {\bf 127}, 140602 (2021)}.
  \bibitem{McRoberts-Bilitewski-Haque-Moessner}
  A.~J.~McRoberts, T.~Bilitewski, M.~Haque, R.~Moessner, 
  \href{https://doi.org/10.1103/PhysRevB.105.L100403}{Phys. Rev. B {\bf 105}, L100403 (2022)}.
  \bibitem{short-paper}
  T.~Veness, K.~Brandner, \emph{Reservoir-induced stabilisation of a periodically driven many-body system} (2022).
  \bibitem{Jin-et-al}
  F.~Jin, T.~Neuhaus, K.~Michielsen, S.~Miyashita, M.~A.~Novotny,
  M.~I.~Katsnelson, H.~De~Raedt,
  \href{http://dx.doi.org/10.1088/1367-2630/15/3/033009}{New J. Phys. {\bf 15}, 033009 (2013)}.
  \bibitem{Krech-Bunker-Landau}
  M.~Krech, A.~Bunker, D.~P.~Landau,
  \href{https://doi.org/10.1016/S0010-4655(98)00009-5}{Comput. Phys. Commun. {\bf 111}, 1  (1998)}.
  \bibitem{Newman-Barkema}
  M.~E.~J.~Newman, G.~T.~Barkema, \emph{Monte Carlo methods in statistical physics}, Clarendon Press (1999).
  \bibitem{Parsons}
  J.~D.~Parsons,
  \href{https://doi.org/10.1103/PhysRevB.16.2311}{Phys. Rev. B {\bf 16}, 2311 (1977)}.
  \bibitem{erdelyi}
  A.~Erd\'{e}lyi, \emph{Asymptotic expansions}, No. 3. Courier Corporation (1956).
  \bibitem{Blanes-Casas-Oteo-Ros}
  S.~Blanes, F.~Casas, J.~A.~Oteo, J.~Ros,
  \href{https://doi.org/10.1016/j.physrep.2008.11.001}{Phys. Rep. {\bf 470}, 151 (2009)}.
  \bibitem{Bender-Orszag}
  C.~M.~Bender, S.~Orszag, \emph{Advanced mathematical methods for scientists and engineers I: Asymptotic methods and perturbation theory} (Vol. 1), Springer Science \& Business Media (1999).
  \bibitem{Berry}
  M.~Berry, \emph{Asymptotics beyond all orders},  Springer, Boston, MA (1991).
  \bibitem{mori-floquet-prethermal}
  T.~Mori,
  \href{https://doi.org/10.1103/PhysRevB.98.104303}{Phys. Rev. B {\bf 98}, 104303 (2018)}.
  \bibitem{feldman}
  E.~B.~Fel'dman,
  \href{https://doi.org/10.1016/0375-9601(84)90027-6}{Phys. Lett. A {\bf 104} (1984)}.
  \bibitem{rubio-abadal}
  A.~Rubio-Abadal, M.~Ippoliti, S.~Hollerith, D.~Wei, J.~Rui, S.~L.~Sondhi,
  V.~Khemani, C.~Gross, I.~Bloch,
  \href{https://doi.org/10.1103/PhysRevX.10.021044}{Phys. Rev. X {\bf 10}, 021044 (2020)}.
  \bibitem{Ma-Dudarev-Semenov-Woo}
  P.-W.~Ma, S.~L.~Dudarev, A.~A.~Semenov, C.~H.~Woo,
  \href{https://doi.org/10.1103/PhysRevE.82.031111}{Phys. Rev. E {\bf 82}, 031111 (2010)}.
  \bibitem{Brown}
  W.~F.~Brown, \href{https://doi.org/10.1103/PhysRev.130.1677}{Phys. Rev. {\bf 130}, 1677 (1963)}.
  \bibitem{Nakajima}
  S.~Nakajima,
  \href{https://doi.org/10.1143/PTP.20.948}{Prog. Theor. Phys. {\bf 20}, 948 (1958)}.
  \bibitem{Zwanzig}
  R.~Zwanzig,
  \href{https://doi.org/10.1063/1.1731409}{J. Chem. Phys. {\bf 33}, 1338 (1960)}.
  \bibitem{Breuer}
  H.-P.~Breuer, F.~Petruccione, \emph{Theory of Open Quantum Systems}, Oxford (2002).
  \bibitem{Breuer-Petruccione}
  H.-P.~Breuer, F.~Petruccione, 
  \href{https://doi.org/10.1103/PhysRevA.55.3101}{Phys. Rev. A {\bf 55}, 3101 (1997)}.
\end{references}
\end{document}